\documentclass[manuscript,screen,authorversion]{acmart}

\AtBeginDocument{%
  \providecommand\BibTeX{{%
    \normalfont B\kern-0.5em{\scshape i\kern-0.25em b}\kern-0.8em\TeX}}}

\usepackage[linesnumbered]{algorithm2e}
\usepackage{amsthm}
\usepackage{subfigure}

\usepackage{xpatch}
\makeatletter
\xpatchcmd{\ps@firstpagestyle}{Manuscript submitted to ACM}{}{\typeout{First patch succeeded}}{\typeout{first patch failed}}
\xpatchcmd{\ps@standardpagestyle}{Manuscript submitted to ACM}{}{\typeout{Second patch succeeded}}{\typeout{Second patch failed}} 
\@ACM@manuscriptfalse
\makeatother
\setcopyright{none}
\renewcommand\footnotetextcopyrightpermission[1]{}
\begin{document}

\title{New Attacks and Defenses for Randomized Caches }

\author{Kartik Ramkrishnan}
\email{ramkr004@umn.edu}
\author{Antonia Zhai}
\email{zhai@umn.edu}
\author{Stephen Mccamant}
\email{mccamant@umn.edu}
\author{Pen Chung Yew}
\email{yew@umn.edu}
\affiliation{%
  \institution{University of Minnesota, Twin Cities}
}

\begin{abstract}

The last level cache is vulnerable to timing based side channel attacks because it is shared by the attacker and the victim processes even if they are located on different cores. These timing attacks evict the victim cache lines using \emph{small conflict groups} (SCG), and monitor the cache to observe when the victim uses these cache lines again. A conflict group is a collection of cache lines which will evict the target cache line.

To defeat these attacks, defenses randomize the address-to-set mappings in hardware, using cryptographic hash functions. Furthermore, these defenses also change the encryption key periodically, to make attacks even harder. The re-randomization rate is slow, because cache lines need to be moved in order to re-randomize, and moving too many cache lines creates extra evictions and affects performance. We show that CEASER needs a substantially higher refresh rate to defend against attacks, resulting in significant performance hit ( 20\% average, upto 50\%). CEASER-S adds another level of randomization to the cache by dividing the cache into smaller banks, known as \emph{skews}, and using a different encryption key for each skew. Thus, each cache line has a different set mappings in each skew. We introduce new attacks on CEASER-S that can learn SCGs in $O(N)$ time. We show that the refresh rate requirement for the default CEASER-S configuration is high to defend these attacks, resulting in significant performance hit (15\%). We also propose to increase the cache associativity and the number of skews for greater security. 

We identify two key issues regarding the previous strategy, CEASER-S, namely the high cost of scaling up the skew count and high cost of increasing refresh rate. Next, we propose a new randomization strategy using an indirection table, which mitigates these two issues. Addresses of cache lines are encrypted and used to lookup the indirection table entry. Each indirection table entry stores a mapping to a randomly chosen cache set. The cache line is placed into this randomly chosen set. The set mappings are re-randomized across \emph{all} sets, which is much greater than the number of skews (100x more than the default configuration), and requires only one or two extra iTable lookups compared to the baseline, thus the cost of randomization is not too bad. Secondly, the encryption key changes upto 50x faster than CEASER's default rate, by using evictions to trigger the re-randomization. Instead of moving cache lines, this mechanism re-randomizes one iTable entry at a time, whenever the cache lines corresponding to the iTable entry are naturally evicted. Thus, the miss rate is not much worse than the baseline.

We quantitatively show that our scheme does almost as good as a fully associative cache to defend against these attacks. We also demonstrate new attacks that target the indirection table by oversubscribing its entries, and quantitatively show that our scheme is resilient against new attacks for trillions of years. 
Using CACTI 7.0, we estimate low area and power overhead compared to a baseline inclusive last-level cache. Lastly, we evaluate its performance overhead using the SPECrate 2017 and PARSEC 3.0 benchmarks, and show that the impact on performance is very low (<4\%).
\end{abstract}

\maketitle

\section{Introduction}
 A side-channel is any means of observation that can indirectly infer secret information about a system or application. Commonly exploited side-channels  include acoustic~\cite{shamir2004acoustic}, electromagnetic~\cite{longo2015soc, gandolfi2001electromagnetic}, and thermal~\cite{hutter2013temperature} side-channels. Side-channel attacks use different components of the processor's microarchitecture to infer secret information, such as speculative execution, branch prediction, TLBs, virtual addressing, and caches~\cite{vanbulck2018foreshadow,weisse2018forshadowNG,lipp2018meltdown,kocher2018spectre,weisse2018forshadowNG,evtyushkin2018branchscope, koruyeh2018spectre,gras2018translation,liu2015last,disselkoen2017prime+,kayaalp2016high}. 
Last-level cache (LLC)  attacks~\cite{liu2015last,kayaalp2016high,bernstein2005cache}  are particularly dangerous because they do not require the victim's and attacker's processes to be co-located on the same core. Cache side-channel attacks can be classified into many different types~\cite{he2017secure,lyu2018survey}. We discuss different kinds of cache side-channel attacks in detail in \S\ref{section:background}.
In particular, conflict-based attacks, such as PRIME + PROBE~\cite{liu2015last} and EVICT + RELOAD~\cite{gruss2015cache}, use interference between the victim and attacker processes to steal secret information. This is achieved using \textit{small conflict groups} (SCG), which are a small group of cache lines which evict the target cache line. We empirically set the maximum size of a small conflict group as 1000 cache lines, because attacks usually require 2000-5000 cycles between measurements~\cite{yarom2014flush+,gruss2016flush+}, and a conflict group of size 1000 takes about 5000 cycles to load, considering the latest LLC implementations~\cite{skylake}, which pipeline the LLC accesses to happen with ~5 cycles between accesses. For a set associative cache, these conflict groups map to the same cache set, thereby known as a \emph{conflict set}, which only needs to be larger than the cache associativity.
In previous work~\cite{liu2016newcache,wang2007new,maas2013phantom,brasser2017dr}, the cache line addresses have been randomized to defeat such attacks. This randomization makes it difficult for the attacker to discover SCGs. These randomization strategies use both hardware~\cite{wang2007new,liu2016newcache,ceaser,qureshi2019new} and software based approaches~\cite{brasser2017software,stefanov2013path}. Software based approaches have large performance overhead (upto 15x) due to introduction of new memory accesses and instructions to randomize the addresses. Hardware-based randomization strategies randomize the mappings between the cache line addresses and the cache sets, and can have much lower overhead (1\%-5\%). \emph{Table based}(TB)~\cite{liu2016newcache,wang2007new} randomization has been proposed for the private cache. TB randomization has a two-level structure. The addresses are mapped to a randomization table in the first level and then the table entries are mapped to the cache sets in the second level. There are two kinds. The first kind, which we call TB-1~\cite{liu2016newcache}, maps addresses to a (conceptually) fully associative randomization table(s) in the first level, and the second level has a static mapping to the cache sets. These fully associative tables are very expensive in terms of energy and area. They can be implemented efficiently for the private caches, but are impractical for the LLC, because it is many times larger. The second kind, which we call TB-2~\cite{wang2007new}, uses multiple randomization tables, one for each process/security domain. In each table, the first level mapping is static, and the second level mapping uses \emph{dynamic random placements} (DRP) i.e. each table entry stores a set mapping, which is randomly changed using a random number generator (RNG), each time the cache line(s) mapped to that entry are evicted. Since the randomization table effectively stores pointers to the cache sets, we call it an \emph{indirection table}, or iTable. It is not suitable for the LLC, where several threads/processes may be active simultaneously, causing the storage overhead to become too high, due to the large number of tables required. \textit{Dynamic encryption}(DE) schemes, such as CEASER~\cite{ceaser} encrypt the cache line address and map the encrypted address to the cache sets. The execution is divided into \textit{epochs}, each of which spans many memory accesses. There is a \textit{current} key and \textit{target} key. At the start of an epoch, all cache sets are mapped using the current key, but they are gradually refreshed so that at the end of the epoch they all use the target key. In order to refresh the cache set, all the cache lines need to be moved to a new location determined by the target key. Therefore, the refresh rate needs to be low (eg. one refresh every 100 accesses), to make the cache line movement overhead low. Dynamic encryption increases security, but is vulnerable to faster attacks, which may find a conflict set in much less time than the epoch, as we show in~\S\ref{section:oldr}.
  
   A general approach to achieve greater security is to add additional levels of defenses. This is a similar approach to multi-factor authentication (MFA), wherein using a password and answering a security question to log in to an online banking account, is significantly more secure than only using a password, but does not take much more login time. The cost the user pays for this extra protection is additive, but the security benefits are multiplicative. CEASER-S~\cite{qureshi2019new} uses this approach i.e. it uses both \emph{random skew select} (RSS) to improve the security of CEASER. It uses a randomized version of the skew associative cache, in which the cache is divided into several smaller caches of lower associativity (called skews) and cache lines may randomly be placed into any of the skews. The second randomization uses the CEASER style dynamic encryption on each skew. However, lower skew counts of CEASER-S are vulnerable to new attacks (see \S\ref{subsub:dedrs}), which can discover SCGs in $O(N)$ time. CEASER-S requires a large number of skews (>128) to defend against these new attacks, or a much higher refresh rate (epoch length $N$ accesses).

 Instead of using skews, we propose to use an indirection table (iTable), \emph{shared} by all processes/threads, to address two key issues, namely, increase the amount of randomization and the refresh rate, at a low cost. The cache line addresses are encrypted and the iTable lookup occurs using the encrypted address. Each iTable entry stores a mapping to a cache set, determined by a random number generator. The cache line is placed in this cache set. Re-randomization occurs one iTable entry at a time (which usually maps one or two cache line addresses). Each re-randomization changes the set mapping to a random value. Thus the re-randomization happens across \emph{all} cache sets, which is 100x greater than the default number of skews of CEASER-S. To increase re-randomization rate at a lower cost, the iTable entry is transitioned to the target key whenever the cache lines mapped by it are naturally evicted, due to cache misses. An epoch length of $2*N$ accesses ($N$ is the cache size) is sufficient to transition most of the iTable entries, which is upto $50x$ more than the randomization rate proposed by CEASER. A cleaner mechanism transitions any iTable entries which did not naturally transition by the end of the epoch.
 
  The above scheme using the indirection table uses both dynamic encrypt in the first level(DE) and dynamic random placements in the second level (DRP), therefore it is called DE+DRP. DE+DRP significantly increases the SCG size, making the cache as secure as a fully associative cache (which has truly random set mappings) against the new attacks. We also introduce a new attack on DE+DRP that targets the indirection table, by \emph{oversubscribing} one of the entries, i.e. create an SCG that maps to a single iTable entry. Using a simple analysis, we show in \S\ref{section:security} that DE+DRP prevents the creation of SCGs for trillions of years.

     In summary, the major contributions of this work are listed below:
    \begin{itemize}
        \item 
        We discuss new attacks that can defeat existing approaches (such as CEASER and CEASER-S) by recovering SCGs in $O(N)$ time.
        \item
        We propose a novel two level approach using an indirection table to defend against conflict-based attacks, called DE+DRP. We show analytically that it is resilient against attacks for trillions of years.
        \item
        We evaluate our scheme for performance, area and energy overhead and find that they are low. The performance evaluation is done on SPEC2017rate and PARSEC 3.0 benchmarks, using ZSim~\cite{sanchez2013zsim}. The area and energy overhead area evaluated using CACTI 7.0~\cite{balasubramonian2017cacti}.
    \end{itemize}
       
       The rest of the paper is organized as follows. \S\ref{section:background} provides background information about existing conflict-based attacks.
       \S\ref{sec:tb_style} discusses existing table based defenses, and why they are not scalable to the LLC.
       \S\ref{section:oldr} discusses existing defenses for the LLC and new attacks that they are vulnerable against.
       \S\ref{sec:tlr} discusses a simple two level randomization with static mappings.
       \S\ref{section:dedrp} discusses the DE+DRP randomization in detail.
       \S\ref{section:security} quantitatively shows that DE+DRP is robust against even new attacks for trillions of years.
       \S\ref{section:misc} discusses miscellaneous design issues.
       \S\ref{section:perf} presents the performance results.
       \S\ref{section:related_work} discusses other related work regarding cache randomization. \S\ref{section:conclusion} concludes the paper.
\section{Attack Background}
\label{section:background}
Cache side-channel attacks come in many different flavours ~\cite{he2017secure}. The majority of cache side-channel attacks use timing information about the victim process, such as total execution time, or the time to access certain addresses in the cache. Three types of cache side-channel attacks have been identified, namely, conflict-based, flush-based, and collision-based. Conflict-based attacks leverage conflict misses in the cache.  Whenever a victim interferes with the attacker's data, or vice-versa, it leaks information about the victim's activities to the attacker. Flush-based attacks use \textit{clflush} or related instructions to flush specific cache lines and monitor their future activities. Lastly, cache collision attacks observe the effect of a victim's cache access patterns on its total execution time to deduce its secret. The key difference between conflict attacks and cache collision attacks is that the attacker does not need to probe the cache directly in the latter, and only needs to measure the execution time of the victim. 

\subsection{Overview of Cache Side-Channel Attacks}
 
A PRIME + PROBE attack~\cite{bernstein2005cache} monopolizes the target set(s) with attacker's data, and waits for a fixed interval. After the interval has elapsed, the attacker measures the time required to access its data. If any cache line was evicted due to a victim access, the attacker would observe greater latency to fetch the cache line, due to a cache miss. Thus, the attacker detects  victim accesses to the target set(s). EVICT + TIME~\cite{bernstein2005cache} attack  evicts victim data from a target set, then executes the victim process and measures the total time of execution. If that set was accessed by the victim, the execution time will be higher due to cache miss(es). This  is used by the attacker to infer secret information used by the victim. EVICT + RELOAD~\cite{gruss2015cache} is very similar to PRIME + PROBE, except that the attacker probes for a \emph{shared} cache line, instead of a cache miss on its data. Lastly, PRIME + ABORT~\cite{disselkoen2017prime+} uses a hardware transaction to prime the target set. If the transaction aborts, it means the attacker data was evicted from the set due to a victim access. The time of access is recorded during each abort, thus revealing the victim access patterns.

FLUSH + RELOAD~\cite{yarom2014flush+} flushes a cache line using \textit{clflush}, waits for a fixed interval, and measures time required to reload the flushed cache line. A higher access latency due to a cache miss indicates that no victim access took place during the previous interval, and vice versa. FLUSH + FLUSH~\cite{gruss2016flush+}  flushes a cache line using \textit{clflush}, waits for a fixed interval, and then flushes the same cache line again. A shorter flush time indicates that no victim access took place during the previous interval, and vice versa. Both these attacks are useful to spy on the victim's access patterns, if the victim and attacker share the same cache line.
\textit{clflush} is also used in speculative execution attacks, such as Spectre~\cite{kocher2018spectre}, Meltdown~\cite{lipp2018meltdown}, and Foreshadow~\cite{vanbulck2018foreshadow,weisse2018forshadowNG} 
to transmit the result of speculative execution to the attacker.

There are two phases in cache collision attacks.
The first phase is a cleaning phase, during which the attacker evicts all the victim data from the cache. In the second phase, the attacker executes the victim process and measures the execution time. Depending on the number of cache hits/misses, the execution time will be different for different inputs. This can be used to deduce the secret information~\cite{bonneau2006cache}.

\subsection{Conflict v\/s Non-Conflict Attacks}
Flush-based attacks, a type of non-conflict based attacks, can be mitigated by disabling data sharing between the victim and the attacker.  
On the other hand, to defend against conflict-based attacks, the application should be \emph{cache oblivious}, i.e. its cache access pattern should be independent of the secret information it tries to protect. In this work, we focus on defense against conflict-based attacks, which are the only mode of attack when the attacker and victim do not share data. We do not defend against flush based attacks, where the attacker is monitoring a cache line shared with the victim. 
 We also do not prevent some covert channel attacks, such as priming the entire cache instead of cache set, detecting cache contention etc. We do prevent shared data attacks which require higher timing resolution, empirically 2000-5000 cycles long~\cite{yarom2014flush+}. However, we do not prevent shared data attacks that might work even without any high resolution timing measurements~\cite{van2017telling}, and leave those to future work.
\section{Table Based Randomization}
\label{sec:tb_style}
Table based randomization is used to prevent conflict based attacks in the L1 cache. Table based randomization's strength is that it re-randomizes the set mapping of a cache line, whenever the cache line is evicted. This high rate of re-randomization makes it difficult to mount attacks. There are two kinds of randomization that are possible for the first level cache, which we name TB-1 and TB-2. In the TB-1 approach, the cache line addresses are mapped to a fully associative randomization table. Thus, each cache line address may be mapped to any entry of the table. The mapping from the table to the cache sets is deterministic. Therefore, in order to access the cache, the entire table needs to be looked up (all entries). This is very expensive for larger caches, like the last level cache. Whenever a cache line is evicted from the cache set, then the next time it is fetched, it is mapped to a random location in the randomization table. Thus the set location of the cache line is also random. This makes attacks that use conflict sets ineffective, because each time the target cache line is evicted, it is simply remapped to a different set. In the TB-2 approach, the cache line addresses are mapped to the randomization table in a deterministic way, using some index bits of the cache line addresses. The randomization table has a pointer in each entry to the correct cache set, where the cache line will be placed. Therefore, we call it the indirection table, or \emph{iTable}. The random number generator determines a cache set which is stored in the iTable entry. There are multiple such iTables, and it is expected that the attacker and victim processes use different iTables. Thus, TB-2 is also not suitable for the LLC because dozens of iTables may be required depending on the core count, making the storage overhead impractical.  
\section{Existing Randomizations and New Attacks}
\label{section:oldr}
In this section, we discuss existing defenses, and new attacks which can defeat the randomization defenses for the LLC. The attacks use small conflict groups (SCGs) (< 1000 cache lines in size) to evict a target cache line. A special case of SCG is a \emph{conflict set}, in which all cache lines of the SCG always map to the same set as the target, and is \emph{guaranteed} to evict the target. The conflict set needs to be at least as large as the cache associativity to evict the target. We use the following variables, $N$ is the number of cache lines, $w$ is the cache associativity.
\subsection{The Set Associative Cache}
The memory address 64 bits. The first few bits (from the right) are the \emph{block offset} bits, which are not used for determining the set mapping. The next few bits are the \emph{set} bits, which determine the set mapping. The remaining bits are the tag bits. For a 64-byte cache line, there are 6 block offset bits ($2^6 = 64$). For a 2 MB LLC cache bank, there are 2048 cache sets, thus there are 11 set bits ($2^{11} = 2048$). We use this configuration for the rest of this work.

\subsubsection{Simple Attack On Set Associative Cache}
\label{subsub:simple}
The simplest attack has two steps. First, we choose a random collection of $L$ cache lines, such that it contains a conflict set (SCG whose cache lines all map to the same set as the target, size at least equals the cache associativity). The cache has $N$ cache lines. The conflict group is of size $L = k*N$, where $k$ is a fraction less than $1$. The next step is to reduce the size of the conflict group in a systematic way so that the attacker is left with a conflict set in the end. 
The simple approach reduces the size one cache line at a time, and checks to see whether the conflict set is still in the reduced conflict group (i.e. there is at least one cache miss when the entire group and the target cache line is loaded). Thus, the total number of accesses to perform the attack is the arithmetic series
\[ L + (L-1) + (L-2) + ...  = O(L^{2}) = O(N^{2})\]
Since the refresh time for DE strategy is $O(N)$, which is much smaller than $O(N^2)$ it is secure for arbitrarily sized caches against such an attack strategy (see~\cite{ceaser})
This simple attack was proposed in a previous work~\cite{liu2015last} to find the conflict sets in the last level cache. 

\subsection{Static Encrypt (SE)}
One approach to protect the cache is by \emph{encrypting} the cache line addresses, using a cryptographic hash function.
We consider the example of a 64-bit address space, and 64-byte cache blocks. 58 address bits except for the first 6 address bits  (from the right) can be randomized by encrypting them. As a result, cache lines are randomly mapped to different sets, because the set bits change to random values. However, the simple attack, which we discuss in \S\ref{subsub:simple}, can still work to create conflict sets.

\subsection{Dynamic Encrypt (DE)}
\label{subsec:de}
Dynamic encryption (DE) uses a dynamically changing cryptographic hash function to encrypt the addresses. These encrypted addresses are then used to map to the cache sets.
This approach was used in CEASER~\cite{ceaser}. The addresses are encrypted using a low latency block cipher , and mapped to the cache sets. Two keys are in use at a time, and the mappings transition gradually between the original key and the target key, during each epoch (a fixed number of cache accesses), by slowly refreshing the cache sets to only use the target key. At the end of the epoch, the current key and the target key are swapped, and the new target key is randomized to a different value. As the refresh routine scans through the cache sets over the course of the epoch, it moves the cache lines in the current set (which uses the original key) to other sets, based on the target hash function. The refresh time linearly depends on the cache size, $N$, and needs to be at least an order of magnitude more than $N$ ($100*N$ accesses default) to ensure that performance is not affected.

Unsophisticated cache attacks require $O(N^{2})$ time to find a conflict set.
However, more sophisticated attacks reduce the attack time to only $O(N)$ time, and are able to break the defense for arbitrarily sized caches. In order to defend against these more advanced attacks, it is required to substantially increase the refresh rate, so that there is a large gap between the attacker's rate of finding the conflict sets and the rate at which the sets are refreshed. Unfortunately, increasing the refresh rate to a safer value (at least an order of magnitude less than $N$) results in a very poor performance, due to a large increase in cache evictions caused by moving cache lines. Also, the hardware cost of refreshing becomes high because, at higher refresh rates, there needs to be hardware support to move multiple cache lines at a time. In the following, we discuss the simple attack, that the DE is able to defend, and more advanced attacks, that can break the defense.

\subsubsection{The Binary Search Attack}
\label{subsub:binary_search}
The simple attack that we discussed in \S\ref{subsub:simple} requires $O(N^2)$ accesses in order to find a conflict set. However, the encryption key of DE gets changed in $O(N)$ time. Therefore, for larger caches, by the time the attacker finds a conflict set, the encryption key would have changed, making the attack ineffective. Instead of reducing the conflict group size one cache line at a time, we propose to reduce it by a small fraction $f$ of its current size. This creates a high probability that the conflict set is retained in the reduced conflict group. The total number of cache accesses is reduced to a geometric series 
\[ L + f\cdot L + f^{2}\cdot L + ...  \leq \frac{L}{1-f} = O(N)\]

\subsubsection{The Advanced Builder Attack}
\label{subsub:advanced_builder}
 Instead of using group reduction, we can build a conflict set one cache line at a time. With each access, the target cache line is also monitored, to check if it has been evicted. If so, the cache line is added to the conflict set. This \emph{builder} attack takes about $N.w$ time, which is $O(N)$, time to discover the conflict set. 

 We can speed up the builder attack in the following manner. First a conflict group is created by randomly selecting $L$ cache lines. The conflict group contains at least one conflict set. When all the cache lines of the conflict group and the target cache line are accessed, there will be at least one cache miss, due to the conflict set inside the conflict group. Next, the conflict group and the target cache line is loaded again, there will be another cache miss, this time on a different cache line of the conflict set, depending on how the replacement state of the cache set got modified due to our accesses. We do this a few times, with a different cache line being revealed due to each cache miss.
Thus, the total attack time is close to $w.L$, which is $O(N)$.

\emph{Impact of Replacement Policy : } There can be a few variations of the builder attack based on the replacement policy. For a cache using the \emph{random} replacement policy, the cache line evicted from the set is randomly selected. In the builder attack, each time a cache line is evicted, a new cache line may be evicted from the cache, revealing a new member of the conflict set, or the same cache line may be evicted again, thus revealing no new member of the conflict set. Therefore, more evictions may be required to find out all members of the conflict set, due to repeat evictions of some of the cache lines. Modern caches typically do not use random replacement, which is a \emph{stateless} replacement policy. Instead, they typically use a \emph{replacement state}, which is often stored using counters for each cache line. These counters are updated whenever the cache set is accessed. An attacker with knowledge of the replacement policy could manipulate these counters to implement a builder style attack faster than for a cache with a random replacement policy. For the rest of the paper, we assume a random replacement policy is used, for maximum protection against attack.

\subsection{Static and Dynamic Random Skew}
The traditional skew associative, which we call static random skew (SRS) cache is vulnerable to attack due to static skew mappings. The simple attack is effective in learning a conflict set for this cache. A more secure variant of the cache uses a random number generator instead of using a deterministic hash function to select the cache skews, which we call dynamic random skew (DRS). This cache is resilient against the usual attacks, such as the builder attack and the conflict group reduction attack. However, it is still possible to develop an SCG using the builder attack.

\subsubsection{Defense Against Basic Attack and Reduction Attack}
In the attack, we create a large conflict group (1000s of cache lines) that contains a conflict set (of size $w+1$) on one of the skews. This will cause a cache miss when the entire conflict group is loaded. When the conflict group is loaded again however, there is only a $1/s$ chance  that the evicted cache line from the conflict set maps to the same skew. If it gets loaded into a different cache skew, then the attack has failed, because no matter how many times the conflict group is loaded, there aren't any further cache misses, hence we don't learn anything about the conflict set for the target cache line. Similarly, the attacks that try to reduce a larger part of the conflict group at a time also fail due to the same reason.

\subsubsection{Builder Attack on DRS cache}
\label{subsub:builder}
A cache line is selected, which is the target cache line. Next, random cache lines are selected, all the while monitoring the target to see whether it is evicted or not. In case it is evicted, then we record the evictor into the SCG. Thus, we build up an SCG in this manner. However, there is a key difference regarding the size of the SCG required to make this attack work. We noted that the minimum size of the SCG was just $w$ for the previous cache designs. However, it can be much larger for the DRS cache depending on the number of skews. Using the analysis below, we find that the minimum skew size to have a significant chance of evicting the target cache line is about $s.w$. We verified using simulations that this SCG size has a significant chance of evicting the target cache lines ( > 50\% ), for $s$ and $w$ values upto 128. The attack time is roughly $N.s.w$, thus linearly increasing it by a factor of $s$ compared to the previous attack. 

\emph{ Analysis : }We perform a simple analysis to gain an intuition for the SCG size required to do an attack. Let us build a conflict group of size $g$. During the process of discovering the members of the SCG, we discover $g/s$ cache lines from each skew, because they are randomly discovered among the skews. Once the members of the SCG are discovered we proceed to use it in an attack. Let the target cache line reside in skew $x$, set $y$. When we load the SCG, then out of the $g/s$ cache lines which map to skew $x$ and set $y$, only $g/s^2$ cache lines will map to skew $x$, set $y$ again, due to random skew selection. We equate this to the set associativity of each skew ($w/s$), which we consider to as enough to evict the target cache line. This yields an SCG size $g=w.s$, which is sufficient to evict the target cache line.

\emph{Measurement Count :} This attack is quite a bit weaker than the conflict set attacks. In order to make a measurement, the SCG needs to be loaded into the cache. For another measurement, a different SCG needs to be used. Thus, the number of high resolution measurements depends on the number of SCGs that the attacker is able to find. However, although it is a weaker attack, it is still significant that the attacker is able to make any measurements at all, hence we consider it to be a security risk that needs to be mitigated. On the other hand, the conflict set attacks are more powerful because the same SCG can be re-used again and again to evict the target cache line. Thus, the attacker can make a large number of measurements (before the epoch runs out).

\subsection{ Dynamic Random Skew + Dynamic Encrypt(DRS+DE)}
\label{subsec:de+rss}
A recent work~\cite{qureshi2019new} combined the DRS scheme and CEASER style dynamic encryption to make the cache more secure. This defense is known as the CEASER-S approach. CEASER-S raises the bar for attackers, because there are now two randomizations instead of one. CEASER-S uses a variant of skew associative cache for the first randomization. 

\subsubsection{Builder Attack on DRS+DE}
\label{subsub:dedrs}
The builder attack can be used on the cache. A target cache line is selected, following which other random cache lines are used to evict it. Each cache line which evicts it is recorded by the attacker and added to the conflict group. Similar to the builder attack on DRS, $g=s.w$ is the minimum size of the SCG for a $> 50\%$ chance to evict the target cache line.
It takes roughly $N$ accesses to discover each member of the SCG, so the attack takes $N.s.w$ time to create ($N$ accesses to discover each member of the SCG).

\emph{Analysis : }
The $g=s.w$ cache lines are discovered in $N.s.w$ time on average. For the attack to be successful, we need to discover these $g$ cache lines in less time than the epoch time. Let $E=k.N$ be the epoch of the cache, where k is an integer $>1$. Therefore $N.w.s = k.N$. This yields $k = w.s$. For the smaller skew counts (2-4) recommended in CEASER-S, and an associativity of 16, $k$ is between $32$ and $64$. Thus, the epoch length needs to be at least $64.N$. We recommend having a significant amount of gap between the attack rate and the refresh rate (at least an order of magnitude), which yields an epoch time of $6.4N$ accesses for the four skew case. We propose that either the epoch length needs to be reduced drastically, or the skew count needs to be increased substantially in order to have a good security guarantee. For a default skew count of 2, we need to have an epoch length of minimum $32.N$ to defend our attack. However, with a smaller SCG size ( $< w.s.N$), probabilistic attacks may yet be possible. We leave such attacks to future work.
  
\emph{Proposed Improvements :} We observe that two improvements are required for our solution. Firstly, it would be much more secure to have a randomization which increased the SCG size by one or two orders of magnitude. If we randomize across all the sets instead of the skews, then our group size would increase to $S.w$ (where $S$ is the number of sets, much greater than the number of skews). Second, a higher rate of change of the encryption key (consequently a lower epoch duration) is beneficial for security too. Using an indirection table, we can achieve both these goals, using the following two strategies. Firstly, the indirection table randomizes across all the cache sets. Secondly, the iTable re-randomizes its entries whenever there is an eviction. The overhead of the iTable is mainly due to storage. The access overhead is quite minimal, requiring only an extra access or two.
\section{Using an Indirection Table For Greater Security}
\label{sec:tlr}
We observe from the previous section that there are two key issues that we need to resolve, namely, (1) we need the randomization to happen across all the sets to create a larger SCG and (2) we need the encryption key to change faster, at a low cost. There can be numerous ways to achieve these goals, so we present one possible solution, using an iTable. Other solutions could include combinations of  skews, indirection tables, using tables in a different way to add randomization, such as using a random offset etc. We provide the following key intuition to do our design. Firstly, in order for the randomization to occur across all the cache sets, we use an indirection table to decide the set mappings. Therefore, the encrypted address does not map to the cache lines directly, and instead maps to the indirection table. Secondly, in order to reduce cost of lookup, each cache line should only lookup a small number of iTable entries to find its set mapping. Lastly, the randomization rate of the encryption key needs to increase significantly to mitigate the attacks, we suggest to re-randomize whenever there are cache evictions to achieve this increase. Another advantage is that there need re-randomization rate when not under attack, i.e. when the cache miss rate is low.

\emph{ Using the iTable for Static Mappings : }
We explain a simplified use of the indirection table first, using static mappings, due to ease of explanation. We consider the first variant of this two level static randomization (TLSR), which we call TLSR-SE, short for TLSR static encrypt. In TLSR-SE, the first level of the mappings are not mapped using the set bits, instead, the cache line address is encrypted first and then mapped to the cache sets. This iTable lookup is called the \emph{select} operation. This makes it harder for the attacker to create the conflict sets, because the attacker does not know the set bits beforehand. The second variant of TLSR is TLSR-SRP, in which the cache line addresses are mapped to the iTable entries using the LSB bits of the address as index bits. The iTable entries contain mappings to the cache sets, which are randomly generated when the cache is initialized. Combining the two above randomization yields TLSR-SE+SRP, i.e. we use both the static encrypt and static random placements, or SE+SRP for short. In the first level, the cache line address is encrypted to map to an entry of the iTable, and in the second level, the iTable entry is mapped to the cache set. This determines the location of the cache line. 
Both TLSR-SE and TLSR-SRP randomization is vulnerable to attack. The key vulnerability is that the mappings are static, hence we can use the basic attack to reduce a conflict group to an SCG, that we discussed in \S\ref{subsub:simple}.

\emph{The iTable Size :}
The iTable size determines how many cache lines can be kept in the cache. To an extreme, only one entry means only one set can be stored. A larger number of entries can store a larger number of cache line. A minimum of $S$ entries is required to populate the entire cache. 
\section{Indirection Table Based Two Level Dynamic Randomization (TLDR) }
\label{section:dedrp}
We observed that the TLSR scheme is not secure due to static mappings in \S\ref{sec:tlr}, because existing attacks can create conflict sets. We consider two simpler variants , which use dynamic encryption (DE) or dynamic random placement (DRP), for improving the security.
Following this, we shall explain the more complex version, i.e. DE+DRP.
For simplicity of explanation, we consider single level dynamic randomization first and then consider multi-level randomization. 

\subsection{TLDR-DE}
We add dynamic encryption to SE+SRP, making it DE+SRP.
There are two key goals that we discussed earlier in \S\ref{subsub:dedrs}, and also need to be incorporated into the scheme.  Using DE, we address the second goal, i.e. we would like to increase the re-randomization rate of the encryption key significantly. There can be a lot of strategies to achieve these goals, so we present just one possible solution, which can be further tuned or may even be totally different, depending on other design goals of the designer. In our solution, we re-randomize one iTable entry at a time, whenever there is an eviction due to a cache miss.

\subsubsection{CEASER-like DE Mechanism Overview}
We use a CEASER-like mechanism to transition the key by slowly transitioning the iTable entries. The execution is divided into epochs, where each epoch is a pre-determined number of cache misses, which we denote as $E$. There are two encryption keys, $k_i$ and $k_j$, in use during an epoch, the current key and the target key, which are used to perform a select operation on the iTable entries, instead of using only a single key that is used by SE encryption. Thus, each address can selects upto two iTable entries, $i$ and $j$, one using the current key, $k_i$ and one using the target key $k_j$. The iTable entries may be in a \emph{transitioned} state, or a \emph{non-transitioned} state. The entries which are non transitioned, can be selected using either key, $k_i$ or $k_j$. The entries which are transitioned can only be selected using the target key, $k_j$. Thus, we may need to check upto two sets $S_i$ and $S_j$ to access the cache. To remove this overhead, we can optimized the key select using a a precedence logic, which is discussed in \S\ref{subsub:key_precedence}. We slowly transition the iTable entries, so that by the end of the epoch, they are all in a transitioned state. We use a transition bit per iTable entry to store this information.
 At this point, we switch the current and target keys, and assign a new randomly chosen target key.
 The iTable entries are changed to the transitioned state, whenever there is an eviction in the cache. This achieves our second design goal of the system. We discuss the details of the transition in the following section. Some optimizations to the replacement policy are also suggested, discussed in \S\ref{subsub:repl}, to facilitate a faster transition rate of the entries, with lower overhead. In this strategy, instead of moving the cache lines when transitioning an iTable entry, we use natural evictions to facilitate the transitions. However, since natural evictions may happen randomly to different iTable entries, it is also necessary to have a cleaner mechanism to ensure that all entries are transitioned. 

\subsubsection{Key Select Operation With Precedence Logic}
\label{subsub:key_precedence}
Ordered key select is an optimization that requires to lookup only one cache set, instead of a couple, thus reducing the access latency. The two iTable entries, determined by the target key, $i$ and $j$, where $i$ has greater precedence than $j$. If $i$ has not yet transitioned to the target state, then it is selected, and we use the set mapping of that entry, $S_i$, and do not need to use $j$. In case $i$ has transitioned to the target state, then it is rejected, and we lookup the set mapping $S_j$ to access the cache line. Thus, we only need to access one of the cache sets, $S_i$ or $S_j$, which is much more efficient than accessing both. 

\subsubsection{Transitioning the iTable entries}
\label{subsub:repl}
Whenever all the cache lines which are mapped by a particular iTable entry, $i$, are evicted, then the iTable entry enters into a transitioned state. If it had already transitioned earlier, then it is not necessary. Sometimes, there may be multiple cache lines which are mapped using the same iTable entry. In this case, we cannot transition until all these cache lines are evicted.  For the purpose of faster transitioning using evictions, we suggest changes to the replacement policy, so that \emph{all} these lines get evicted during a cache miss, not just one cache line. Furthermore, since the transitions happen in a random manner depending on the evictions, it is also necessary to have a cleaner mechanism to transition iTable entries which may have been missed.  

\emph{Replacement Policy For Faster Transitions : } The cache set has many cache lines. Per cache line, we suggest to keep track of the iTable entry corresponding to the cache line address (to reduce encryption overhead). When there is a cache miss on cache line $L$ with corresponding iTable entry $k$, then the replacement policy does two things. Firstly, it classifies the cache lines according to their respective iTable entries, selects a random iTable entry among these, and evicts the corresponding cache lines. 

\emph{Cleaner Mechanism :} The cleaner mechanism requires to transition the iTable entries so that all entries have transitioned to the target encryption key at the end of the epoch. The cleaner mechanism's main goal is to ensure correctness, because any cache lines which did not transition may result in an incorrect lookup. The cleaner mechanism activates in the second half of the epoch, and scans through the iTable entries. Whenever it detects an iTable entry that has not transitioned, it initiates an eviction in order to transition the iTable entry to the transitioned state. In most cases, the iTable entry would already have transitioned, so there is not much perfomance impact. 

\emph{The iTable Size and the Epoch Length : } Since we may evict multiple cache lines per cache miss, therefore, we recommend to have a low iTable load, so that there is usually only 1 cache line mapped by a single iTable entry. Thus an iTable size of $N$ (average load of 1) or greater is recommended.
The epoch length depends on the iTable size. Since we would like most of the iTable entries to have naturally transitioned by the end of the epoch (for lower cleaner mechanism overhead), we use an epoch of length $2*N$.

\emph{Total Storage Overhead :} We are trading off storage overhead and security i.e. we have increased security but pay for it with storage overhead. The iTable contains $N$ entries, and each entry has $log2(S)$ bits for the set mapping, and 1 bit to indicate the transition state. The other overhead is the per cache line iTable entry information in the cache sets, which is $log2(N)$ bits per cache, and the increased tag size (since the tag now includes all bits except the set bits). The overhead is about $2*log2(S)+1+log2(N)$ bits per cache line, which is about 368 bits for $S=2^{11}$ and $N=2^{15}$. Since each cache line has about 550 bits, this is about 7\% overhead.

\begin{algorithm}[]
 \label{alg:dedrp} 

  \tcc{Cache and iTable Storage. N is the number of cache lines, S is the number of cache sets, and w is set associativity}
  iTable[N]\{
  
   ~~ refresh; \tcp{refresh bit}
   ~~ set\_mapping; \tcp{set mapping}
  \}
  
  cache[S][w];
  
  \tcc{Pseudocode for the DE part}
  
  check\_buffer(lAddr, buf\_hit, data)\;
  
  \If {buf\_hit} { 
  
  return data\;}

  \tcc{Key Select Operation uses current key DE1() and target key DE2()}
  iIdx1 = DE1(lAddr); \tcp{get iTable entry iIdx1}
  
  iIdx2 = DE2(lAddr); \tcp{get iTable entry iIdx2}
  
  r1 = iTable[iIdx1].refresh; \tcp{get refresh bit r1 }
  
  S1 = iTable[iIdx1].set\_mapping; \tcp{get set mapping S1}
  
  r2 = iTable[iIdx2].refresh; \tcp{get refresh bit r2 }
  
  S2 = iTable[iIdx2].set\_mapping; \tcp{get set mapping S2 }
 
  \tcc{Check cache hit in set indicated by iTable}
  \eIf{r1}{ 
  
     check\_tags(cache, S1, tag\_hit, data) \tcp*{access S1}
  }{
  
     check\_tags(cache, S2, tag\_hit, data) \tcp*{access S2}
      
  }
  
  \tcc{Pseudocode for the DRP part}
  \eIf{tag\_hit}{
  
     return data   \tcp*{tag hit, we are done}
  }{
  
     data = access\_mem( lAddr ) \tcp*{tag miss, get data from memory}
     
     repl(cache, S, repl\_ways, oversubscribe) \tcp*{invoke replacement policy calculation}
     
     \eIf{oversubscribe}{
     
         use\_buffer(data, lAddr)  \tcp*{use buffer to store oversubscribed cache line}
         
     }{
     
         refr\_idx = evict(cache, repl\_ways)  \tcp*{Evict the cache line(S) selected by replacement policy repl(), return the corresponding iTable entry refr\_idx}
         fill\_set(cache, data, S, lAddr) \tcp*{Put the data into the cache set}
         
         refresh(iTable, refr\_idx) \tcp*{refresh the iTable entry}
     }
  }
 \caption{Pseudocode for DE+DRP}
\end{algorithm}

\subsubsection{Attack on TLDR-DE}
The attack on TLDR-DE requires to find a conflict set before the DE randomization can change. The attack can be carried out using the advanced attack strategies that we discussed in \S\ref{subsub:advanced_builder}, where the conflict set can be reduced very quickly, even before the encryption function can change. We consider one particular example for convenience. In this example, we create a conflict group. Each time there is a cache miss, we learn a new member of the conflict set inside the conflict group. Thus in $k.N$ cache accesses, it is possible to learn all the cache lines of the conflict set. The epoch length is about $2*N$ cache accesses, therefore, it is not long enough to prevent this attack, which happens in $L*w$ evictions, which can be less than $2*N$. For example, we verified using simulation that we recovered a conflict set for $N=32768$ and $w=16$ (typical LLC configuration) in $0.4*N$ accesses.

\subsection{TLDR-DRP}
We consider a second variant of TLDR, which we call SE+DRP. In this variant we re-randomize the second level mappings between the iTable entries and the cache sets, so that set mapping of a cache line is randomly changed after each eviction. This achieves the second goal that we discussed in \S\ref{subsub:dedrs}. 

\subsubsection{The DRP Mechanism}
The basic structure of the TLDR-DRP contains an iTable which creates two levels of mappings, one between the cache line addresses and the iTable entries, and the other layer between the iTable entries and the cache sets. The iTable entry contains the set mapping. Whenever the cache lines corresponding to the iTable entry are evicted, then the re-randomization occurs in the iTable entry. Whenever the iTable entry transitions, then we change the set mapping stored in the iTable entry. 

\emph{ Defending Against Oversubscription Attacks :}
An extra component we add is a \emph{victim buffer}, which holds the cache lines which may be used in an \emph{oversubscription} attack on the cache. Whenever an entry of the iTable is oversubscribed, then the extra cache lines are sent into the victim buffer. This helps increase the attack time until the buffer has become full. The larger the buffer size, the greater the time to do the attacks. We discuss these oversubscription attacks in more detail in the following.

\subsubsection{Attack on TLDR-DRP}
\label{subsub:oversub}
It is possible to attack the TLDR-DRP scheme because the first level randomization is static. We call this attack the \emph{oversubscription} attack. The key idea is to load a conflict group, where all cache lines map to the same iTable entry as the target cache line. This means that these cache lines will \emph{always} map to the same cache set as the target. We can gradually build up the SCG. We load the target cache lines and a random cache line. Then, we load random cache lines, monitoring if both the target $L$ and the candidate $C$ are evicted simultaneously. If so, then C is added to the SCG. We repeat until the SCG grows to size $w$.

\subsection{TLDR-DE+DRP}
We use both DE and DRP mechanisms previously discussed in order to achieve both the goals of randomization that we stated in \S\ref{subsub:dedrs}.
Since there are many components now, we use a pseudocode, to explain this more complex scheme.

\subsubsection{PseudoCode}
We present the pseudocode for DE+DRP in Algorithm 1, which contains all the mechanism we have discussed so far. When the cache is accessed, then the first step is to access the \emph{victim buffer} ( lines 6-9). This victim buffer is required to prevent the attacks on DE, as we discussed in \S\ref{subsub:oversub}. If the cache line is not in the victim buffer, then the next step is to check in the cache. In order to access the cache, the first step is the optimized key select mechanism that we discussed in \S\ref{subsub:key_precedence} (lines 10-15). The key select decides which iTable entry needs to be accessed. Once we determine this, we access the cache set mapped by the iTable entry (lines 16-20). Next, we check the tags in the cache set for a tag hit (lines 16-20). If the cache set contains the tag, then we are done. Otherwise, it is a tag miss, hence the memory is accessed and the cache line is fetched from main memory (lines 24). Next, the replacement policy that we discussed in \S\ref{subsub:repl} decides which cache lines needs to be evicted (line 25). Once the eviction is done, then the cache line is filled either into the cache set or into the victim buffer, based on whether the cache set was oversubscribed or not (26-31). Lastly, the corresponding iTable entry is refreshed (line 31). We have omitted the pseudo code for the cleaner mechanism to keep our description short. There can be many ways to implement it, the way we did it was to have an extra access by the cleaner after each normal access during the second half of the epoch.

\section{Security Analysis}
\label{section:security}
The attack criteria for the attack to succeed is to form an SCG to evict the target cache line. We try all the attacks on the previous schemes and show that they do not work on DE+DRP. The first kind of attack is the basic attack (see \S\ref{subsub:simple}), which tries to reduce the conflict group to an SCG. Each time there is a cache eviction, then the target cache line is re-randomized to a different cache set, and only a $1/S$ chance to map back to the same cache set. It is virtually impossible to reduce the conflict group beyond a few cache lines.

  The builder attack (\S\ref{subsub:builder}) will learn one cache line at a time, and add it to the SCG. However, after each eviction, the target cache line will map to a random cache set. Thus, the conflict group needs to become comparatively larger in order to attack the cache. We can analyze how many cache lines are needed to evict the target cache line. Intuitively, each member of the SCG will map to a random set, therefore, it will require many cache lines to evict the target cache line. Using simulation. Therefore, our scheme significantly increases the security of the cache compared to the baseline. 

Using simulation, we recorded the total time to evict the targets using SCGs of different sizes. We found that we need at about 1000 cache lines for a 2\% chance to evict the target cache lines, and for a large probability like $50\%$, it will require an even greater SCG size, closer to $N$ cache lines. Thus, our strategy increases the size of the conflict group by an order of magnitude compared to the existing strategy, which only requires about 100 cache lines for a $70\%$ chance to evict the target in the default configuration (2 skews), and requires a larger number of skews (about 64) to get it down to $2\%$. This is much larger than the size required to mount the attack.

\emph{Analysis :}
To simplify the analysis, we model the conflict group as a random set of cache lines, of size $g$. The cache configuration is also randomly chosen. Assuming that the target cache line is in a random cache location, the SCG has approximately $1/N$ chance to evict it. Thus, the chance to evict the target cache line survives after each eviction is about $(1-1/N)$. As we load the entire SCG, the chance that the target cache line decreases exponentially, depending on the SCG size, to $(1-1/N)^g$. Substituting in the group size of about $N$ yields about $1/e$ chance to evict the target. If we cut down the SCG size to about $1000$ cache lines, then the probability to evict the target falls to 3\%. These numbers are in agreement with our simulation results. The analysis we have performed is exactly the same as what we may do for a fully associative cache, which has truly random set mappings. 

\subsection{Oversubscription Attack and Defense for the iTable}
Instead of directly attacking the cache, the attacker can target the iTable entries instead, by \emph{oversubscribing} the iTable entries. This is an attack similar to the one described in \S\ref{subsub:oversub}, where we tried to oversubscribe one of the iTable entries in order to attack the cache. In order to carry out this attack, the attacker tries to learn iTable entries that map to the same set as the target cache line. First, the attacker loads a conflict group, which contains an SCG oversubscribing one of the iTable entries. Next the attacker tries to reduce the size of the conflict group one cache line at a time in order to learn the SCG.  Therefore, in order to prevent this type of attacks, our epoch needs to complete fast enough, so that the SCG gets scattered before the attacker can discover it. We found using simulation that an epoch of length $2*N$ is enough to prevent the iTable entries from being oversubscribed completely.

\emph{ Analysis :}
We estimate the probability that a particular iTable entry is oversubscribed given a particular epoch length. The accesses to each entry of the iTable is modelled as a Poisson distribution in which cache lines are evenly spread over all the iTable entries. The Poisson Distribution is described using the following description :
\[ e^{-{\lambda}}\lambda^{-k}/k! \]
Where $\lambda$ is the load. The load on the iTable entries refers to the number of cache lines which are mapped by the iTable entries. The total iTable load is quite low, because only a small number of cache lines are mapped to each iTable entry. The epoch length is $2*N$ cache lines, therefore, we consider the average load on an entry using $2*N$ cache lines (2), for our calculations.  We set $\lambda=2$ and the $k=9$, in order to oversubscribe the iTable entry (for cache set associativity $w=8$), and multiply the resultant probability with the number of iTable entries, to find out how many iTable entries might have $9$ mapped cache lines, which is about 16. Since all these cache lines are stored inside of the victim buffer, it is impossible to oversubscribe a single entry of the iTable. Using simulations, we observed at most 11 cache lines oversubscribed per epoch for the benchmarks. Therefore, we use a victim buffer size of 32 (2x) to guarantee that we always have room for cache lines which oversubscribe the iTable entry.
 
\section{Miscellaneous Discussion}
\label{section:misc}
  We discuss some miscellaneous issues for a real implementation of the scheme, which we organize into three categories, the LLC pipeline, the replacement policy implementation and other system level aspects.
  
  \emph{Pipelining : } 
  Modern caches are pipelined for high performance. The LLC pipeline needs an extra stage in order to integrate the iTable into it.
  Thus, the total access latency of the cache increases by one pipeline stage. Using modern processors as a reference~\cite{skylake,broadwell}, this requires about 3-7 cycles of extra access time.
  In the common case, the iTable stage rarely causes pipeline hazards. Thus we conservatively model the delay using a 10 cycle additional latency, which would allows 5 cycles for the hash function and 5 cycles for the iTable pipeline stage. Even shorter delay is possible using lower latency hash function or more optimized design.

\emph{Replacement Policy : } The replacement policy needs to check the iTable index of each cache line (stored per cache line) and select one of them for eviction. This is similar to other high performance replacement schemes~\cite{wu2011ship, jaleel2010high}, except that they check a smaller number of bits. It is possible to efficiently implement this, by only checking the last few LSB bits for example and checking the remaining bits if necessary. The storage overhead of the iTable index bits required for replacement decision is discussed in ~\S\ref{subsec:area}.

 \emph{Other System Level Aspects : } Our design is independent from the rest of the system and does not require much changes. There is no change in the coherence protocol, including multi-socket coherence, or the way memory consistency is handled by the cache. Context switches and TLB shootdowns do not require participation from the cache because it is the LLC, which is physically tagged and indexed. There is no change required in SMT or core design. Two level randomization can even be implemented for directories, which have a similar set based structure as caches. Hardware/Software prefetchers require no modification.

\begin{figure*}[t!]
 \begin{subfigure}{} \includegraphics[trim=0 90 0 0,clip,width=\linewidth]{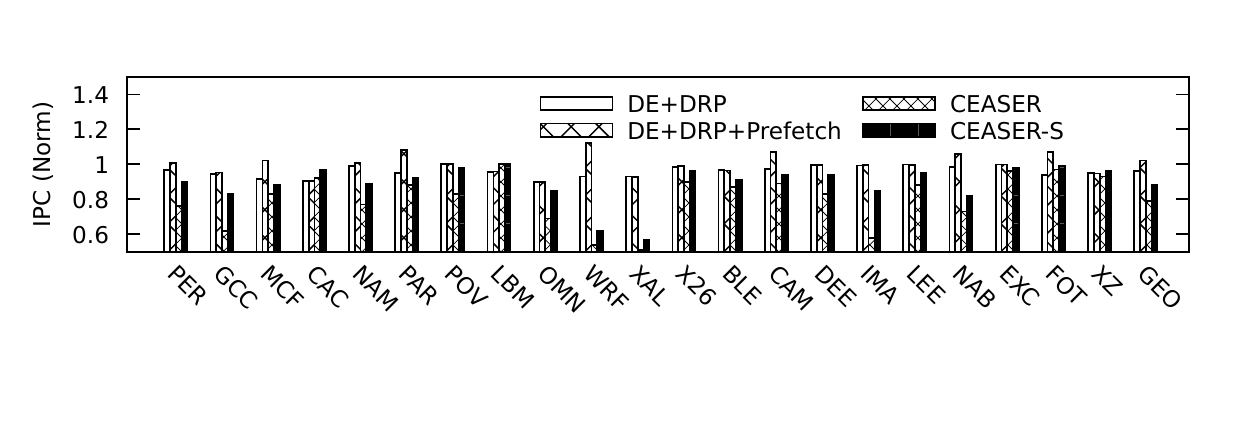}
 \label{fig:spec_one_core_ipc}
  \end{subfigure}
\vspace*{-2.7cm}

 \begin{subfigure}{} \includegraphics[trim=0 90 0 0,clip,width=\linewidth]{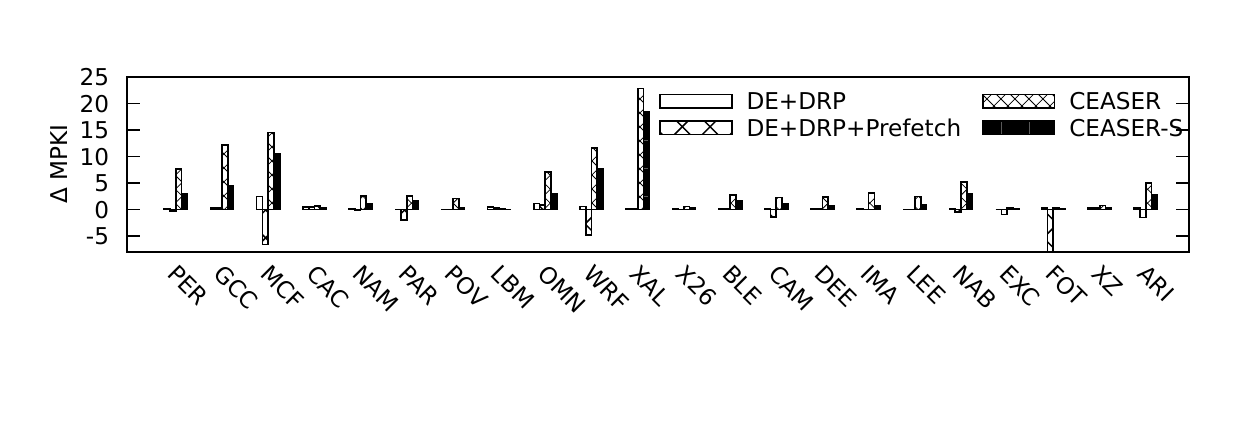}
\label{fig:spec_one_core_mpki}
  \end{subfigure}
\vspace*{-1.9cm}

\begin{subfigure}{}
\includegraphics[trim=0 85 0 20,clip,width=\linewidth]{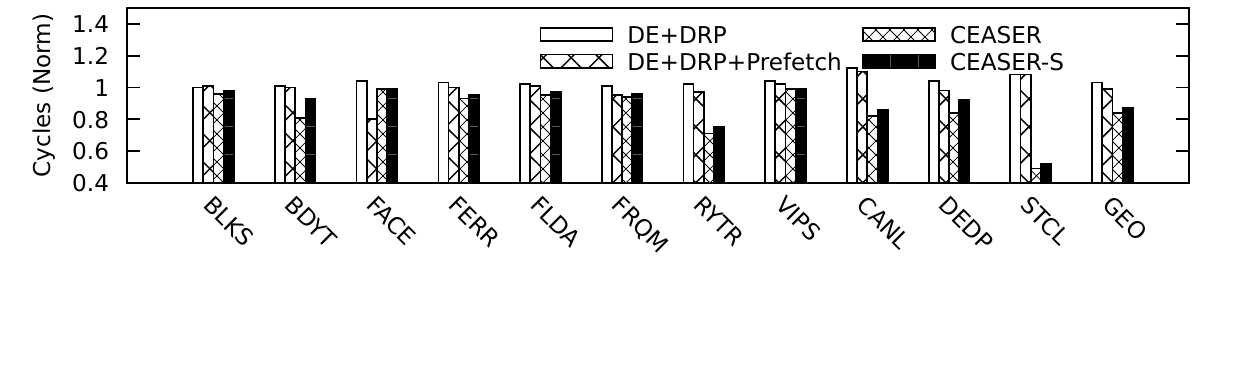}
\label{fig:PARSEC_performance}
\end{subfigure}
\vspace*{-2.3cm}

\begin{subfigure}{}
\includegraphics[trim=0 85 0 20,clip,width=\linewidth]{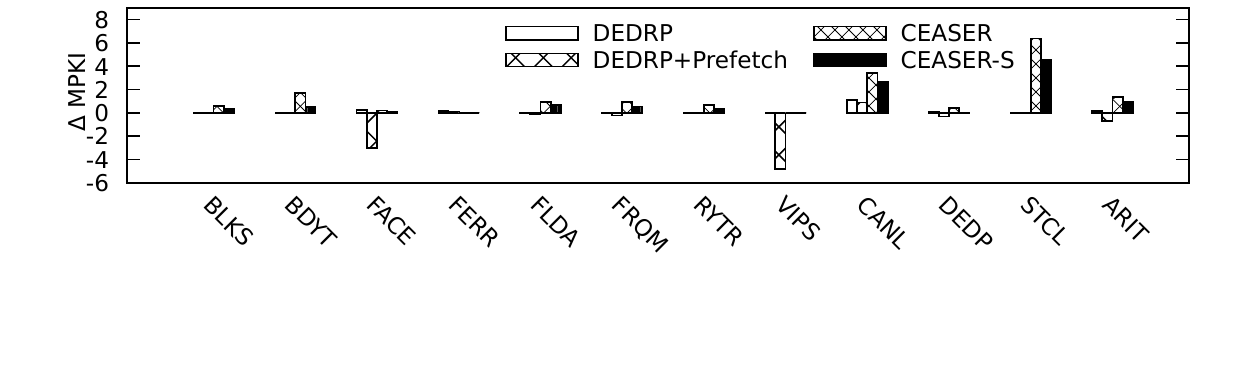}
\label{fig:PARSEC_mpki}
\end{subfigure}
\vspace*{-2.4cm}

\begin{subfigure}{}
\includegraphics[trim=0 90 0 10,clip,width=\linewidth]{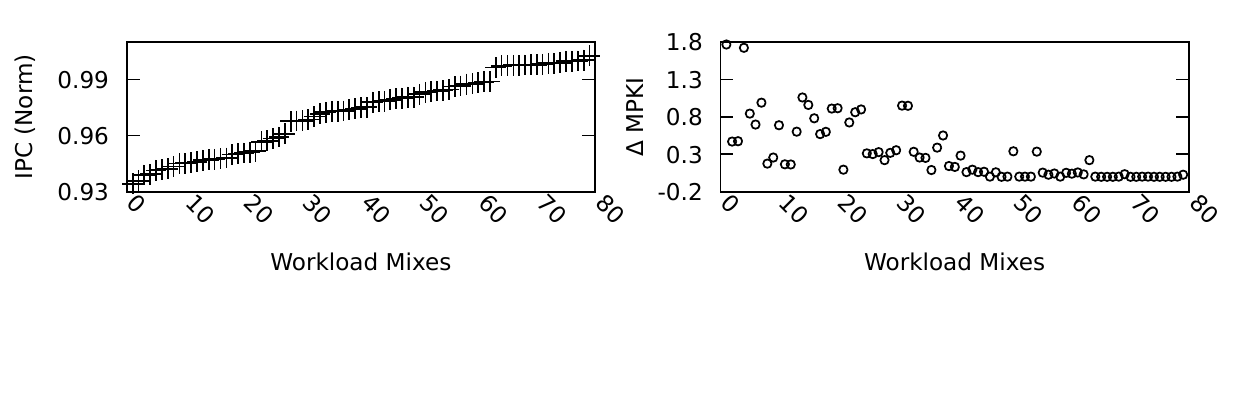}
\label{fig:spec_mixes_mpki}
\end{subfigure} 
\vspace*{-2.2cm}
\label{fig:performance}
\caption{ From Top To Bottom, SPECRate 2017 IPC, PARSEC 3.0, MultiProgramming IPC and MPKI}
\end{figure*} 

\section{Evaluation}
\label{section:perf}
We simulated the SPECrate 2017 and PARSEC 3.0 benchmarks using five configurations : (1) the baseline LLC, which is an inclusive set associative cache (2) DE+DRP, and (3) DE+DRP with prefetching enabled (DE+DRP+pref) (4) CEASER (5) CEASER-S. The results for the latter four configurations are normalized against the baseline. The LLC bank has the same configuration as the example we used throughout the paper, i.e. an iTable size of $2^{15}$, and an epoch size of $2^{16}$ evictions. To account for the additional latency of the iTable access and the encryption, we added 10 cycles to the baseline LLC access time to model the randomized LLC access time (see \S\ref{section:misc}).
The two key results we study are the instructions per cycle and misses per kilo instructions (MPKI). We  study total execution time instead of IPC for parsec benchmarks, because they have variable number of instructions. The IPC is normalized by dividing the baseline IPC. The $\Delta$ MPKI is obtained by subtracting the MPKI of baseline from the MPKI of baseline.

\subsection{Single Core Performance Results}
Each SPECrate benchmark is simulated for a representative interval~\cite{sherwood2001basic} of 1 billion instructions using the `ref' inputs, whereas the PARSEC benchmarks are simulated in their entirety using the `simmedium' inputs.
We simulate 21/23 SPECrate benchmarks and 11/13 PARSEC 3.0 benchmarks. We had runtime errors with \texttt{roms}, \texttt{bwaves} and do not have results for these benchmarks. Each simulation was performed five times (due to the random nature of the access patterns), and the average result is presented.
The simulated configuration uses an out-of-order processor   32KB L1 cache, 256KB L2 cache and 2MB L3 cache banks per core. The prefetcher is a strided prefetcher that trains on L2 accesses and prefetches into L2 and L3. 

\subsubsection{CEASER and CEASER-S :}
We simulated the performance for CEASER using an epoch length 0.1*N accesses. This was required in order to mitigate the fastest attack, which found a conflict set in less than N accesses, using the strategy in \S\ref{subsub:binary_search}. The performance fell by over 20\% on average for SPEC2017 rate benchmarks. The maximum decrease in performance was for \texttt{gcc}, \texttt{mcf}, \texttt{wrf} and \texttt{xalan}, due to the large increase in the MPKI. Some bencmarks had a significant drop in performance even if there wasn't a large increase in the MPKI, this may be because the relative increase in MPKI is very large, even if the absolute value is not that large. For example, the MPKI increase for \texttt{imagick} is only about 4, but the absolute MPKI for the baseline was only about 0.017, thus the relative MPKI increase was close to 200x.

\emph{CEASER-S :}
A key configuration is to evaluate CEASER-S on the baseline 16-way associative cache, where the cache was divided into two skews. We set the epoch length to $N$ accesses, which is ten times slower randomization rate than CEASER, thus we expect the performance to be better. The performance penalty was still about 15\% for the SPEC bencmarks. The maximum penalty we observed was for \texttt{xalan} and \texttt{wrf}, due to a large increase in the MPKI. The performance was mostly in line with CEASER, but the effect on MPKI was less severe, thus the performance was not as bad. 

\subsubsection{DE+DRP}
The performance results normalized against the baseline inclusive cache configuration, and shown in  Figure 1. There was a 4\% IPC penalty for the randomized cache on average, and a marginal increase in MPKI ($<$ 1\%) for the SPECrate 2017 benchmark suite on a single-core processor.  We observe that benchmarks \texttt{mcf},\texttt{wrf} and \texttt{fotonik} benefit significantly from prefetching, due to a large decrease in MPKI (5, 3 and 8 respectively). The other benchmarks have only a small increase in the MPKI. The primary reason for lower performance is due to the increased latency of accessing the randomized LLC. \texttt{xalan} shows the maximum degradation in performance. However, \texttt{xalan} does not show significant increase in MPKI, only about 0.24. The significant performance drops is due to a high rate of LLC accesses; about 10 \% of all loads occur in the LLC. Thus, the higher access latency has a significant impact on the performance regardless of the MPKI, so it is worth it to design the system with lower LLC latency.

\emph{DE+DRP+Prefetch : } This configuration observed a small increase in performance on average compared to the baseline, due to less number of conflict misses.  With prefetching enabled, the performance became 2\% better than the baseline, on average. The maximum benefit was observed for fotonik ( > 10\% performance benefit).

\subsection{SPECrate 2017 Multiprogramming Performance}    
 We classified the benchmarks into high MPKI(\texttt{mcf}, \texttt{lbm}, \texttt{parest}, \texttt{cam4}, \texttt{bwaves}), medium MPKI(\texttt{xz}, \texttt{deepsjeng}, \texttt{cactus}, \texttt{x264}, \texttt{gcc}, \texttt{omnetpp}, \texttt{namd}) and low MPKI(\texttt{pov},  \texttt{exchange}, \texttt{blender}, \texttt{leela}, \texttt{wrf}, \texttt{imagick}, \texttt{fotonik}) groups.
 There are four groups of 20 workloads each using benchmarks with different MPKI. 
 MIX-1 has only high MPKI benchmarks. MIX-2 has both high MPKI and low MPKI benchmarks, whereas MIX-3 has only low MPKI benchmarks. We simulated the baseline and the DE+DRP configuration for 80 different workload mixes. The bottom subfigure of Figure~1 shows the IPC and MPKI of various workload mixes compared to the baseline inclusive cache, for a four-core configuration, ordered in increasing performance. We fast forwarded 10 billion cycles, then simulated until both benchmarks had reached 1 billion cycles.  We used the weighted IPC metric to measure the performance. Most of the benchmarks show a small decrease in IPC; the performance reduction was about 3\% on average. 
The increase in MPKI is quite minimal for almost all the workloads, which explains the modest increase in IPC; the average MPKI increase is less than 0.1. Some of the workloads (such as workload 20) have a very small increase in MPKI, yet a larger decrease in performance. This may be because of the performance penalty due to a greater proportion of loads being serviced by the LLC. 

We also simulated 8-benchmark workload mixes on 8-core processors, where the workloads were created using a similar strategy as for the 4-core simulation. We observe about 3\% performance loss on average, similar to the four-core simulation result. 
This is because the random nature of LLC access patterns does not create extra conflict misses when a larger number of processes run concurrently. For the same reason, we expect the randomization scheme to scale well to a larger number of cores.

\subsection{PARSEC 3.0 Multithreaded Performance Results}
We simulated 11/13 parsec benchmarks for all four configurations. Due to runtime errors, we do not have results for \texttt{swaptions} and \texttt{x264}. We simulated a four-core configuration, and ran the complete benchmark using the \emph{simmedium} input, for all four configurations. 

\emph{CEASER and CEASER-S :} For CEASER, the average performance hit was about 20\% and upto 60\% in the case of \texttt{streamcluster}. This was due to significant increase in the MPKI of all the benchmarks.
 For CEASER-S, the performance hit was a bit lower, about 15\% on average, due to the lower rate of re-randomization compared to CEASER. The greatest performance impact was also for \texttt{streamcluster}, with a 55\% performance hit. The main reason was the large increase in MPKI for \texttt{streamcluster}. Some benchmarks, such as raytrace, also had significant performance hit, but the $\Delta$ MPKI was not that much. However, the relative increase in MPKI was quite high for \texttt{raytrace}, about 60x increase in the MPKI.

\emph{ DE+DRP :} The effect of randomization on performance is again minimal for most benchmarks. On average, we did not observe a significant difference in execution time or total cache misses. 
The most performance degradation was about 8\% for \texttt{canneal} and 7\% for \texttt{streamcluster}. This is because both of these benchmarks had a high percentage of loads serviced by the LLC (about 6\% and 5\% each), thus the extra LLC access latency penalty affected the performance. 

\emph{ DE+DRP+Prefetch :} On average, there was a small performance benefit due to prefetching into LLC. \texttt{facesim} showed a significant reduction in runtime with prefetching enabled because there was a 3x reduction in LLC load misses. 
Since the other benchmarks had a much lower MPKI with almost all of them being less than 1, there wasn't much change in the overall MPKI and the LLC accesses were a relatively less important factor for performance compared to some of the SPEC benchmarks like MCF and LBM, which have much higher MPKI. 8-core and 16-core configurations also showed similar performance as baseline. PARSEC benchmarks also showed about 15\% decrease in performance, mainly due to a significant increase in the MPKI.

\subsection{Area and Storage Overhead}
\label{subsec:area}
We used CACTI 7.0 for modelling the area increase in the cache, and also calculated the storage increase for each scheme. CEASER and CEASER-S did not have much storage overhead. We also did not model the area overhead of the encryption circuits, which are only thousands of gates in size, and require a negligible amount of extra area.

\emph{CEASER :} CEASER requires a significantly higher refresh rate, with an epoch length of $0.1*N$. Therefore, we suggest to use 10 extra read-write ports to handle the extra traffic. There is significant area impact of this due to the extra bitlines required to read the storage cells. Using CACTI 7.0, we estimate the area to increase by 30x, which is not practical.

\emph{CEASER-S :} There are two key changes in the hardware. Firstly, there are two skews in the cache, which requires some duplication of the set lookup logic. Therefore, we expect to have a small increase in area. We did not consider this area increase, because we expect it is not that much. Secondly, we add one extra read-write port to handle the extra cache line movement, which results in area increase of about 2x.

\emph{DE+DRP : } Compared to the baseline cache, the key sources of storage overhead are the iTable (64 KB), the extra tag bits in the tag array (12 extra tag bits per cache line), and the iTable index bits used by the replacement policy (16 bits per cache line). The tag size originally was 48 bits, and the cache lines are 64 bytes each. The total storage overhead is about 7\%. The area of the iTable, when implemented as a direct-mapped cache, is very small compared to the tag and data arrays. The tag array sizes increase significantly compared to the baseline, due to the extra index bits and larger tag size. However, it is still very small compared to the data array's size.
 Finally, the data array size is significantly smaller when the associativity is decreased. The randomized cache array's size is about 10\% smaller than the baseline due to lower associativity.
Using CACTI 7.0, we estimated that the total area of the tag and data arrays reduced by 2\% for an 8-way associative randomized cache, compared to a 16-way associative baseline inclusive cache. This is because the increase in area due to the three sources of overhead was offset by the reduction in associativity.

\subsection{Energy Consumption}
We used CACTI 7.0 to estimate the energy cost of each cache access.

\emph{CEASER and CEASER-S :} The energy cost of CEASER increases due to the extra encryption and decryption.
CEASER-S has some additional energy overhead due to the duplicated set lookup logic for each skew, however, we expect this is quite small compared to the tag and data lookup. For both these schemes, the increase in energy occurs due to the larger number of cache accesses (due to cache misses) and longer runtime, and not due to the cost of each cache access. For CEASER-S, we expect about 10x dynamic energy cost due to the extra refresh rate, and about 2x energy cost due to lower refresh rate compared to CEASER.

\emph{DE+DRP :}There are two key sources of dynamic energy overhead. Firstly, there are two iTable access per cache hit, and an additional iTable access per cache miss. Secondly, each access requires to read a larger tag compared to the set associative cache. Fortunately, these overheads can be completely negated by the reduced associativity of the randomized cache compared to the baseline. The randomized cache is 8-way associative, whereas the baseline is 16-way associative. Thus the former has less dynamic energy consumption when accessing the cache and tag arrays.  
 
On average, for the single-core SPECrate benchmarks, we estimate approximately 4\% dynamic energy reduction. For the four-core PARSEC 3.0 benchmarks, we estimate 6\% reduction in dynamic energy. The leakage energy consumption is increases modestly (< 4\%) due to extra tag bits and iTable storage.
For CEASER and CEASER-S, there can be significant increase in dynamic energy due to the performance loss, which results in a significantly longer execution time. In particular, there are a larger number of cache accesses, which results in more energy consumption from the cache as well. Leakage energy is not affected, because both these schemes do not require additional storage. This is a key advantage of these schemes from the perspective of energy consumption.

\subsection{Other Factors that Affect Performance of DE+DRP}
\emph{Hash Function Latency :}
   In our simulations, we imposed 10 cycles additional latency for each LLC access (5 cycles for iTable lookup and 5 cycles for encryption), but it is possible to reduce this latency substantially by removing the encryption from the critical access path, or using a low latency cipher.
   There was negligible degradation in performance compared to the baseline for zero latency, and a much smaller performance penalty for 5 cycle latency or less $(< 1\%)$.

\emph{Associativity :} 
  For associativity 4 or more, we did not observe significant degradation in performance. The 8-way associative cache, which we use in C1, hits a sweet spot in terms of hardware overhead, security and performance.  
  
  \emph{Cleaner Routine Overhead :} 
  The cleaner routine is responsible for carrying out cache line evictions over the course of the last $N$ accesses of an epoch. The performance effect of the cleaner routine was to increase the contention due to extra accesses to the cache and to create some extra evictions. This did not have a significant impact on performance, because in the common case, the cleaner routine access does not cause an eviction in the cache, and only has an extra iTable access.
  
  \emph{Victim Buffer : }
  The victim buffer was filled upto a max of 11 entries, which is well below its 32 entry capacity. Thus, the victim buffer works as expected. Using CACTI 7.0, we estimate negligible increase in power ($<1\%$) due to small buffer size and infrequent data accesses.
\section{Related Work}
\label{section:related_work}
Many software and hardware approaches to defend against cache side-channel attacks exist. The most relevant to our work are \textit{randomization}-based defenses.

CEASER~\cite{ceaser} uses an OLDR-DE approach to defend the cache. CEASER-S~\cite{qureshi2019new}, uses a skew associative cache in addition to OLDR-DE to improve the security over CEASER.

NewCache~\cite{liu2016newcache} uses a CAM address decoder to implement a fully associative randomization table, which is very expensive in terms of dynamic power (10x), leakage power (50\% extra), and area (6x) for a 2MB LLC bank. It gets worse for larger cache sizes. Our approach is more scalable regardless of the number of concurrent processes operating at the time. Multi-table solutions using OLDR-DRP approach~\cite{wang2007new} require different randomization tables for different security domains, which is not scalable to the LLC due to the large storage (many MBs per bank). The single table version of~\cite{wang2007new} is not secure, as we discussed in~\S\ref{section:oldr}, due to the oversubscription attack. In contrast, DE+DRP has the advantage that it is both scalable and secure against attacks.

Path ORAM~\cite{stefanov2013path} creates a large set of memory accesses corresponding to each memory access to hide the true access from the attacker, and has been implemented in hardware~\cite{maas2013phantom}.  Random-eviction caches randomly evict data from cache sets to add noises to the attacker's measurements~\cite{keramidas2008non}, whereas random-fill caches~\cite{liu2014random} de-correlate the demand fetches from the addresses fetched into the cache. However, they do not provide any security guarantee against conflict-based attacks. 

\subsubsection{Software Randomization Techniques}
Software-based randomization~\cite{brasser2017dr} changes the addresses randomly and probabilistically during runtime. ASLR randomizes the virtual addresses used by the kernel (KASLR) and user space~\cite{aslr}. SGX Shield implements ASLR for secure SGX enclaves~\cite{seo2017sgx}. However, ASLR only randomizes the code locations once when they are loaded, so access patterns may still be learned via the cache side-channel over the course of execution. Dr SGX~\cite{brasser2017dr} probabilistically randomizes the addresses using a software `permutation' cache. Memory trace oblivious execution algorithms~\cite{liu2013memory} ensure that the same memory trace is generated regardless of the secret, but incur 15x or more performance overhead, making them impractical for most applications~\cite{liughostrider,rane2015raccoon,liu2013memory}.

\subsubsection{Cache Partitioning}
  While cache randomization is effective to defeat cache side channels, partitioning is even more secure because it causes isolation between the attacker and the victim, ensuring no information leakage, and thus even preventing covert channels. However, partitioning usually is non-scalable, because sharing of resources is no longer based on need. Having too many partitions drastically reduces the system performance.

SecDCP~\cite{wang2016secdcp}, DAWG~\cite{kirianskydawg}, PLCache~\cite{wang2007new} use way partitioning to allocate different ways of a set to different security domains. Since there is never any sharing and interference between these ways, there is no possibility of doing an attack. A related work is SHARP~\cite{yan2017secure}, which uses the cache replacement policy to protect the contents of the private caches from cross-core attacks on the last level cache. NoMo~\cite{domnitser2012non} cache ensures that an attacker cannot monopolize a cache set and thus limits the observation of the victim's cache accesses. CATalyst~\cite{liu2016catalyst} leverages Intel CAT technology to allocate `secure pages' to virtual machines, which get an exclusive access to some of the ways of each cache set. StealthMem uses page coloring~\cite{kim2012stealthmem} to partition the cache sets so that secure and insecure processes cannot interfere with each other. CacheBar~\cite{zhou2016software} uses software partitioning between security domains to disable cache line sharing in the last level cache.

\subsubsection{Miscellaneous}
Hardware performance counters can be used to detect unusual cache activity, as used by HexPads~\cite{payer2016hexpads}.   CC-Hunter~\cite{chen2014cc} looks for strange access patterns in the microarchitecture to detect attacks.
Replay Confusion~\cite{yan2016replayconfusion} and SHARP~\cite{yan2017secure} can detect attacks by checking unusual cache access patterns.
Noises can be added to the cache timing to disrupt the attacker's measurements~\cite{hu1992reducing}, or by changing the granularity of the timers~\cite{martin2012timewarp}.
Some cache side-channel defenses are designed specifically for defeating speculative execution attacks include SafeSpec~\cite{khasawneh2018safespec} and InvisiSpec~\cite{yan2018invisispec}, by bypassing the cache until speculative loads are safe to be committed.

\section{Conclusion}
\label{section:conclusion}
In this work, we have
two key contribution. Firstly, we present new attacks which defeat existing randomization based defeneses which protect the cache.
Secondly, we present a novel two-level randomization scheme for caches, which defeats side-channel attacks by randomizing the set mappings of the cache lines. A mathematical analysis is presented to show that conflict-based attacks will become impractical to carry out. We also proposed a novel design for the last-level cache that implements this scheme. We simulate real-world PRIME+PROBE attacks on AES and RSA ciphers, and show that our randomization is effective.
We used ZSim, a modern multicore simulator, and used SPEC 2017 and PARSEC 3.0 benchmarks to evaluate the performance overhead. We used CACTI 7.0 to estimate the area overhead of our design. Our implementation substantially increases the security of the cache against side-channel attacks, but with very little area and performance overhead. 

\bibliographystyle{ACM-Reference-Format}
\bibliography{sample-journal}


\begin{thebibliography}{58}


\ifx \showCODEN    \undefined \def \showCODEN     #1{\unskip}     \fi
\ifx \showDOI      \undefined \def \showDOI       #1{#1}\fi
\ifx \showISBNx    \undefined \def \showISBNx     #1{\unskip}     \fi
\ifx \showISBNxiii \undefined \def \showISBNxiii  #1{\unskip}     \fi
\ifx \showISSN     \undefined \def \showISSN      #1{\unskip}     \fi
\ifx \showLCCN     \undefined \def \showLCCN      #1{\unskip}     \fi
\ifx \shownote     \undefined \def \shownote      #1{#1}          \fi
\ifx \showarticletitle \undefined \def \showarticletitle #1{#1}   \fi
\ifx \showURL      \undefined \def \showURL       {\relax}        \fi
\providecommand\bibfield[2]{#2}
\providecommand\bibinfo[2]{#2}
\providecommand\natexlab[1]{#1}
\providecommand\showeprint[2][]{arXiv:#2}

\bibitem[\protect\citeauthoryear{??}{asl}{[n.d.]}]%
        {aslr}
 \bibinfo{year}{[n.d.]}\natexlab{}.
\newblock \bibinfo{title}{{\url{
  https://en.wikipedia.org/wiki/Address_space_layout_randomization } }}.
\newblock
\newblock


\bibitem[\protect\citeauthoryear{??}{bro}{[n.d.]}]%
        {broadwell}
 \bibinfo{year}{[n.d.]}\natexlab{}.
\newblock \bibinfo{title}{https://www.7-cpu.com/cpu/Broadwell.html}.
\newblock
\newblock


\bibitem[\protect\citeauthoryear{??}{sky}{[n.d.]}]%
        {skylake}
 \bibinfo{year}{[n.d.]}\natexlab{}.
\newblock \bibinfo{title}{\url{ https://www.7-cpu.com/cpu/Skylake.html }}.
\newblock
\newblock


\bibitem[\protect\citeauthoryear{Balasubramonian, Kahng, Muralimanohar,
  Shafiee, and Srinivas}{Balasubramonian et~al\mbox{.}}{2017}]%
        {balasubramonian2017cacti}
\bibfield{author}{\bibinfo{person}{Rajeev Balasubramonian},
  \bibinfo{person}{Andrew~B Kahng}, \bibinfo{person}{Naveen Muralimanohar},
  \bibinfo{person}{Ali Shafiee}, {and} \bibinfo{person}{Vaishnav Srinivas}.}
  \bibinfo{year}{2017}\natexlab{}.
\newblock \showarticletitle{CACTI 7: New tools for interconnect exploration in
  innovative off-chip memories}.
\newblock \bibinfo{journal}{\emph{ACM Transactions on Architecture and Code
  Optimization (TACO)}} \bibinfo{volume}{14}, \bibinfo{number}{2}
  (\bibinfo{year}{2017}), \bibinfo{pages}{14}.
\newblock


\bibitem[\protect\citeauthoryear{Bernstein}{Bernstein}{2005}]%
        {bernstein2005cache}
\bibfield{author}{\bibinfo{person}{Daniel~J Bernstein}.}
  \bibinfo{year}{2005}\natexlab{}.
\newblock \showarticletitle{Cache-timing attacks on AES}.
\newblock  (\bibinfo{year}{2005}).
\newblock


\bibitem[\protect\citeauthoryear{Bonneau and Mironov}{Bonneau and
  Mironov}{2006}]%
        {bonneau2006cache}
\bibfield{author}{\bibinfo{person}{Joseph Bonneau} {and} \bibinfo{person}{Ilya
  Mironov}.} \bibinfo{year}{2006}\natexlab{}.
\newblock \showarticletitle{Cache-collision timing attacks against AES}. In
  \bibinfo{booktitle}{\emph{International Workshop on Cryptographic Hardware
  and Embedded Systems}}. Springer, \bibinfo{pages}{201--215}.
\newblock


\bibitem[\protect\citeauthoryear{Brasser, Capkun, Dmitrienko, Frassetto,
  Kostiainen, M{\"u}ller, and Sadeghi}{Brasser et~al\mbox{.}}{2017}]%
        {brasser2017dr}
\bibfield{author}{\bibinfo{person}{Ferdinand Brasser}, \bibinfo{person}{Srdjan
  Capkun}, \bibinfo{person}{Alexandra Dmitrienko}, \bibinfo{person}{Tommaso
  Frassetto}, \bibinfo{person}{Kari Kostiainen}, \bibinfo{person}{Urs
  M{\"u}ller}, {and} \bibinfo{person}{Ahmad-Reza Sadeghi}.}
  \bibinfo{year}{2017}\natexlab{}.
\newblock \showarticletitle{DR. SGX: Hardening SGX Enclaves against Cache
  Attacks with Data Location Randomization}.
\newblock \bibinfo{journal}{\emph{arXiv preprint arXiv:1709.09917}}
  (\bibinfo{year}{2017}).
\newblock


\bibitem[\protect\citeauthoryear{Brasser, M{\"u}ller, Dmitrienko, Kostiainen,
  Capkun, and Sadeghi}{Brasser et~al\mbox{.}}{[n.d.]}]%
        {brasser2017software}
\bibfield{author}{\bibinfo{person}{Ferdinand Brasser}, \bibinfo{person}{Urs
  M{\"u}ller}, \bibinfo{person}{Alexandra Dmitrienko}, \bibinfo{person}{Kari
  Kostiainen}, \bibinfo{person}{Srdjan Capkun}, {and}
  \bibinfo{person}{Ahmad-Reza Sadeghi}.} \bibinfo{year}{[n.d.]}\natexlab{}.
\newblock \showarticletitle{Software grand exposure: SGX cache attacks are
  practical}.
\newblock  (\bibinfo{year}{[n.\,d.]}).
\newblock


\bibitem[\protect\citeauthoryear{Chen and Venkataramani}{Chen and
  Venkataramani}{2014}]%
        {chen2014cc}
\bibfield{author}{\bibinfo{person}{Jie Chen} {and} \bibinfo{person}{Guru
  Venkataramani}.} \bibinfo{year}{2014}\natexlab{}.
\newblock \showarticletitle{Cc-hunter: Uncovering covert timing channels on
  shared processor hardware}. In \bibinfo{booktitle}{\emph{Microarchitecture
  (MICRO), 2014 47th Annual IEEE/ACM International Symposium on}}. IEEE,
  \bibinfo{pages}{216--228}.
\newblock


\bibitem[\protect\citeauthoryear{Disselkoen, Kohlbrenner, Porter, and
  Tullsen}{Disselkoen et~al\mbox{.}}{2017}]%
        {disselkoen2017prime+}
\bibfield{author}{\bibinfo{person}{Craig Disselkoen}, \bibinfo{person}{David
  Kohlbrenner}, \bibinfo{person}{Leo Porter}, {and} \bibinfo{person}{Dean
  Tullsen}.} \bibinfo{year}{2017}\natexlab{}.
\newblock \showarticletitle{Prime+ abort: A timer-free high-precision l3 cache
  attack using intel TSX}. In \bibinfo{booktitle}{\emph{26th USENIX Security
  Symposium (USENIX Security 17),(Vancouver, BC)}}. \bibinfo{pages}{51--67}.
\newblock


\bibitem[\protect\citeauthoryear{Domnitser, Jaleel, Loew, Abu-Ghazaleh, and
  Ponomarev}{Domnitser et~al\mbox{.}}{2012}]%
        {domnitser2012non}
\bibfield{author}{\bibinfo{person}{Leonid Domnitser}, \bibinfo{person}{Aamer
  Jaleel}, \bibinfo{person}{Jason Loew}, \bibinfo{person}{Nael Abu-Ghazaleh},
  {and} \bibinfo{person}{Dmitry Ponomarev}.} \bibinfo{year}{2012}\natexlab{}.
\newblock \showarticletitle{Non-monopolizable caches: Low-complexity mitigation
  of cache side channel attacks}.
\newblock \bibinfo{journal}{\emph{ACM Transactions on Architecture and Code
  Optimization (TACO)}} \bibinfo{volume}{8}, \bibinfo{number}{4}
  (\bibinfo{year}{2012}), \bibinfo{pages}{35}.
\newblock


\bibitem[\protect\citeauthoryear{Evtyushkin, Riley, Abu-Ghazaleh, Ponomarev,
  et~al\mbox{.}}{Evtyushkin et~al\mbox{.}}{2018}]%
        {evtyushkin2018branchscope}
\bibfield{author}{\bibinfo{person}{Dmitry Evtyushkin}, \bibinfo{person}{Ryan
  Riley}, \bibinfo{person}{Nael~CSE Abu-Ghazaleh}, \bibinfo{person}{Dmitry
  Ponomarev}, {et~al\mbox{.}}} \bibinfo{year}{2018}\natexlab{}.
\newblock \showarticletitle{BranchScope: A New Side-Channel Attack on
  Directional Branch Predictor}. In \bibinfo{booktitle}{\emph{Proceedings of
  the Twenty-Third International Conference on Architectural Support for
  Programming Languages and Operating Systems}}. ACM,
  \bibinfo{pages}{693--707}.
\newblock


\bibitem[\protect\citeauthoryear{Gandolfi, Mourtel, and Olivier}{Gandolfi
  et~al\mbox{.}}{2001}]%
        {gandolfi2001electromagnetic}
\bibfield{author}{\bibinfo{person}{Karine Gandolfi},
  \bibinfo{person}{Christophe Mourtel}, {and} \bibinfo{person}{Francis
  Olivier}.} \bibinfo{year}{2001}\natexlab{}.
\newblock \showarticletitle{Electromagnetic analysis: Concrete results}. In
  \bibinfo{booktitle}{\emph{International Workshop on Cryptographic Hardware
  and Embedded Systems}}. Springer, \bibinfo{pages}{251--261}.
\newblock


\bibitem[\protect\citeauthoryear{Gras, Razavi, Bos, and Giuffrida}{Gras
  et~al\mbox{.}}{[n.d.]}]%
        {gras2018translation}
\bibfield{author}{\bibinfo{person}{Ben Gras}, \bibinfo{person}{Kaveh Razavi},
  \bibinfo{person}{Herbert Bos}, {and} \bibinfo{person}{Cristiano Giuffrida}.}
  \bibinfo{year}{[n.d.]}\natexlab{}.
\newblock \showarticletitle{Translation Leak-aside Buffer: Defeating Cache
  Side-channel Protections with TLB Attacks}.
\newblock


\bibitem[\protect\citeauthoryear{Gruss, Maurice, Wagner, and Mangard}{Gruss
  et~al\mbox{.}}{2016}]%
        {gruss2016flush+}
\bibfield{author}{\bibinfo{person}{Daniel Gruss},
  \bibinfo{person}{Cl{\'e}mentine Maurice}, \bibinfo{person}{Klaus Wagner},
  {and} \bibinfo{person}{Stefan Mangard}.} \bibinfo{year}{2016}\natexlab{}.
\newblock \showarticletitle{Flush+ Flush: a fast and stealthy cache attack}. In
  \bibinfo{booktitle}{\emph{International Conference on Detection of Intrusions
  and Malware, and Vulnerability Assessment}}. Springer,
  \bibinfo{pages}{279--299}.
\newblock


\bibitem[\protect\citeauthoryear{Gruss, Spreitzer, and Mangard}{Gruss
  et~al\mbox{.}}{2015}]%
        {gruss2015cache}
\bibfield{author}{\bibinfo{person}{Daniel Gruss}, \bibinfo{person}{Raphael
  Spreitzer}, {and} \bibinfo{person}{Stefan Mangard}.}
  \bibinfo{year}{2015}\natexlab{}.
\newblock \showarticletitle{Cache Template Attacks: Automating Attacks on
  Inclusive Last-Level Caches.}. In \bibinfo{booktitle}{\emph{USENIX Security
  Symposium}}. \bibinfo{pages}{897--912}.
\newblock


\bibitem[\protect\citeauthoryear{He and Lee}{He and Lee}{2017}]%
        {he2017secure}
\bibfield{author}{\bibinfo{person}{Zecheng He} {and} \bibinfo{person}{Ruby~B
  Lee}.} \bibinfo{year}{2017}\natexlab{}.
\newblock \showarticletitle{How secure is your cache against side-channel
  attacks?}. In \bibinfo{booktitle}{\emph{Proceedings of the 50th Annual
  IEEE/ACM International Symposium on Microarchitecture}}. ACM,
  \bibinfo{pages}{341--353}.
\newblock


\bibitem[\protect\citeauthoryear{Hu}{Hu}{1992}]%
        {hu1992reducing}
\bibfield{author}{\bibinfo{person}{Wei-Ming Hu}.}
  \bibinfo{year}{1992}\natexlab{}.
\newblock \showarticletitle{Reducing timing channels with fuzzy time}.
\newblock \bibinfo{journal}{\emph{Journal of computer security}}
  \bibinfo{volume}{1}, \bibinfo{number}{3-4} (\bibinfo{year}{1992}),
  \bibinfo{pages}{233--254}.
\newblock


\bibitem[\protect\citeauthoryear{Hutter and Schmidt}{Hutter and
  Schmidt}{2013}]%
        {hutter2013temperature}
\bibfield{author}{\bibinfo{person}{Michael Hutter} {and}
  \bibinfo{person}{J{\"o}rn-Marc Schmidt}.} \bibinfo{year}{2013}\natexlab{}.
\newblock \showarticletitle{The temperature side channel and heating fault
  attacks}. In \bibinfo{booktitle}{\emph{International Conference on Smart Card
  Research and Advanced Applications}}. Springer, \bibinfo{pages}{219--235}.
\newblock


\bibitem[\protect\citeauthoryear{Jaleel, Theobald, Steely~Jr, and Emer}{Jaleel
  et~al\mbox{.}}{2010}]%
        {jaleel2010high}
\bibfield{author}{\bibinfo{person}{Aamer Jaleel}, \bibinfo{person}{Kevin~B
  Theobald}, \bibinfo{person}{Simon~C Steely~Jr}, {and} \bibinfo{person}{Joel
  Emer}.} \bibinfo{year}{2010}\natexlab{}.
\newblock \showarticletitle{High performance cache replacement using
  re-reference interval prediction (RRIP)}. In \bibinfo{booktitle}{\emph{ACM
  SIGARCH Computer Architecture News}}, Vol.~\bibinfo{volume}{38}. ACM,
  \bibinfo{pages}{60--71}.
\newblock


\bibitem[\protect\citeauthoryear{Kayaalp, Abu-Ghazaleh, Ponomarev, and
  Jaleel}{Kayaalp et~al\mbox{.}}{2016}]%
        {kayaalp2016high}
\bibfield{author}{\bibinfo{person}{Mehmet Kayaalp}, \bibinfo{person}{Nael
  Abu-Ghazaleh}, \bibinfo{person}{Dmitry Ponomarev}, {and}
  \bibinfo{person}{Aamer Jaleel}.} \bibinfo{year}{2016}\natexlab{}.
\newblock \showarticletitle{A high-resolution side-channel attack on last-level
  cache}. In \bibinfo{booktitle}{\emph{Proceedings of the 53rd Annual Design
  Automation Conference}}. ACM, \bibinfo{pages}{72}.
\newblock


\bibitem[\protect\citeauthoryear{Keramidas, Antonopoulos, Serpanos, and
  Kaxiras}{Keramidas et~al\mbox{.}}{2008}]%
        {keramidas2008non}
\bibfield{author}{\bibinfo{person}{Georgios Keramidas},
  \bibinfo{person}{Alexandros Antonopoulos}, \bibinfo{person}{Dimitrios~N
  Serpanos}, {and} \bibinfo{person}{Stefanos Kaxiras}.}
  \bibinfo{year}{2008}\natexlab{}.
\newblock \showarticletitle{Non deterministic caches: A simple and effective
  defense against side channel attacks}.
\newblock \bibinfo{journal}{\emph{Design Automation for Embedded Systems}}
  \bibinfo{volume}{12}, \bibinfo{number}{3} (\bibinfo{year}{2008}),
  \bibinfo{pages}{221--230}.
\newblock


\bibitem[\protect\citeauthoryear{Khasawneh, Koruyeh, Song, Evtyushkin,
  Ponomarev, and Abu-Ghazaleh}{Khasawneh et~al\mbox{.}}{2018}]%
        {khasawneh2018safespec}
\bibfield{author}{\bibinfo{person}{Khaled~N Khasawneh},
  \bibinfo{person}{Esmaeil~Mohammadian Koruyeh}, \bibinfo{person}{Chengyu
  Song}, \bibinfo{person}{Dmitry Evtyushkin}, \bibinfo{person}{Dmitry
  Ponomarev}, {and} \bibinfo{person}{Nael Abu-Ghazaleh}.}
  \bibinfo{year}{2018}\natexlab{}.
\newblock \showarticletitle{SafeSpec: Banishing the Spectre of a Meltdown with
  Leakage-Free Speculation}.
\newblock \bibinfo{journal}{\emph{arXiv preprint arXiv:1806.05179}}
  (\bibinfo{year}{2018}).
\newblock


\bibitem[\protect\citeauthoryear{Kim, Peinado, and Mainar-Ruiz}{Kim
  et~al\mbox{.}}{2012}]%
        {kim2012stealthmem}
\bibfield{author}{\bibinfo{person}{Taesoo Kim}, \bibinfo{person}{Marcus
  Peinado}, {and} \bibinfo{person}{Gloria Mainar-Ruiz}.}
  \bibinfo{year}{2012}\natexlab{}.
\newblock \showarticletitle{STEALTHMEM: System-Level Protection Against
  Cache-Based Side Channel Attacks in the Cloud.}. In
  \bibinfo{booktitle}{\emph{USENIX Security symposium}}.
  \bibinfo{pages}{189--204}.
\newblock


\bibitem[\protect\citeauthoryear{Kiriansky, Lebedev, Amarasinghe, Devadas, and
  Emer}{Kiriansky et~al\mbox{.}}{[n.d.]}]%
        {kirianskydawg}
\bibfield{author}{\bibinfo{person}{Vladimir Kiriansky}, \bibinfo{person}{Ilia
  Lebedev}, \bibinfo{person}{Saman Amarasinghe}, \bibinfo{person}{Srinivas
  Devadas}, {and} \bibinfo{person}{Joel Emer}.}
  \bibinfo{year}{[n.d.]}\natexlab{}.
\newblock \showarticletitle{DAWG: A Defense Against Cache Timing Attacks in
  Speculative Execution Processors}.
\newblock  (\bibinfo{year}{[n.\,d.]}).
\newblock


\bibitem[\protect\citeauthoryear{Kocher, Genkin, Gruss, Haas, Hamburg, Lipp,
  Mangard, Prescher, Schwarz, and Yarom}{Kocher et~al\mbox{.}}{2018}]%
        {kocher2018spectre}
\bibfield{author}{\bibinfo{person}{Paul Kocher}, \bibinfo{person}{Daniel
  Genkin}, \bibinfo{person}{Daniel Gruss}, \bibinfo{person}{Werner Haas},
  \bibinfo{person}{Mike Hamburg}, \bibinfo{person}{Moritz Lipp},
  \bibinfo{person}{Stefan Mangard}, \bibinfo{person}{Thomas Prescher},
  \bibinfo{person}{Michael Schwarz}, {and} \bibinfo{person}{Yuval Yarom}.}
  \bibinfo{year}{2018}\natexlab{}.
\newblock \showarticletitle{Spectre Attacks: Exploiting Speculative Execution}.
\newblock \bibinfo{journal}{\emph{arXiv preprint arXiv:1801.01203}}
  (\bibinfo{year}{2018}).
\newblock


\bibitem[\protect\citeauthoryear{Koruyeh, Khasawneh, Song, and
  Abu-Ghazaleh}{Koruyeh et~al\mbox{.}}{2018}]%
        {koruyeh2018spectre}
\bibfield{author}{\bibinfo{person}{Esmaeil~Mohammadian Koruyeh},
  \bibinfo{person}{Khaled Khasawneh}, \bibinfo{person}{Chengyu Song}, {and}
  \bibinfo{person}{Nael Abu-Ghazaleh}.} \bibinfo{year}{2018}\natexlab{}.
\newblock \showarticletitle{Spectre Returns! Speculation Attacks using the
  Return Stack Buffer}. In \bibinfo{booktitle}{\emph{12th $\{$USENIX$\}$
  Workshop on Offensive Technologies ($\{$WOOT$\}$ 18)}}. $\{$USENIX$\}$
  Association.
\newblock


\bibitem[\protect\citeauthoryear{Lipp, Schwarz, Gruss, Prescher, Haas, Mangard,
  Kocher, Genkin, Yarom, and Hamburg}{Lipp et~al\mbox{.}}{2018}]%
        {lipp2018meltdown}
\bibfield{author}{\bibinfo{person}{Moritz Lipp}, \bibinfo{person}{Michael
  Schwarz}, \bibinfo{person}{Daniel Gruss}, \bibinfo{person}{Thomas Prescher},
  \bibinfo{person}{Werner Haas}, \bibinfo{person}{Stefan Mangard},
  \bibinfo{person}{Paul Kocher}, \bibinfo{person}{Daniel Genkin},
  \bibinfo{person}{Yuval Yarom}, {and} \bibinfo{person}{Mike Hamburg}.}
  \bibinfo{year}{2018}\natexlab{}.
\newblock \showarticletitle{Meltdown}.
\newblock \bibinfo{journal}{\emph{arXiv preprint arXiv:1801.01207}}
  (\bibinfo{year}{2018}).
\newblock


\bibitem[\protect\citeauthoryear{Liu, Harris, Maas, Hicks, Tiwari, and Shi}{Liu
  et~al\mbox{.}}{[n.d.]}]%
        {liughostrider}
\bibfield{author}{\bibinfo{person}{Chang Liu}, \bibinfo{person}{Austin Harris},
  \bibinfo{person}{Martin Maas}, \bibinfo{person}{Michael Hicks},
  \bibinfo{person}{Mohit Tiwari}, {and} \bibinfo{person}{Elaine Shi}.}
  \bibinfo{year}{[n.d.]}\natexlab{}.
\newblock \showarticletitle{GhostRider: A Hardware-Software System for Memory
  Trace Oblivious Computation}.
\newblock  (\bibinfo{year}{[n.\,d.]}).
\newblock


\bibitem[\protect\citeauthoryear{Liu, Hicks, and Shi}{Liu
  et~al\mbox{.}}{2013}]%
        {liu2013memory}
\bibfield{author}{\bibinfo{person}{Chang Liu}, \bibinfo{person}{Michael Hicks},
  {and} \bibinfo{person}{Elaine Shi}.} \bibinfo{year}{2013}\natexlab{}.
\newblock \showarticletitle{Memory trace oblivious program execution}. In
  \bibinfo{booktitle}{\emph{Computer Security Foundations Symposium (CSF), 2013
  IEEE 26th}}. IEEE, \bibinfo{pages}{51--65}.
\newblock


\bibitem[\protect\citeauthoryear{Liu, Ge, Yarom, Mckeen, Rozas, Heiser, and
  Lee}{Liu et~al\mbox{.}}{2016a}]%
        {liu2016catalyst}
\bibfield{author}{\bibinfo{person}{Fangfei Liu}, \bibinfo{person}{Qian Ge},
  \bibinfo{person}{Yuval Yarom}, \bibinfo{person}{Frank Mckeen},
  \bibinfo{person}{Carlos Rozas}, \bibinfo{person}{Gernot Heiser}, {and}
  \bibinfo{person}{Ruby~B Lee}.} \bibinfo{year}{2016}\natexlab{a}.
\newblock \showarticletitle{Catalyst: Defeating last-level cache side channel
  attacks in cloud computing}. In \bibinfo{booktitle}{\emph{High Performance
  Computer Architecture (HPCA), 2016 IEEE International Symposium on}}. IEEE,
  \bibinfo{pages}{406--418}.
\newblock


\bibitem[\protect\citeauthoryear{Liu and Lee}{Liu and Lee}{2014}]%
        {liu2014random}
\bibfield{author}{\bibinfo{person}{Fangfei Liu} {and} \bibinfo{person}{Ruby~B
  Lee}.} \bibinfo{year}{2014}\natexlab{}.
\newblock \showarticletitle{Random fill cache architecture}. In
  \bibinfo{booktitle}{\emph{Microarchitecture (MICRO), 2014 47th Annual
  IEEE/ACM International Symposium on}}. IEEE, \bibinfo{pages}{203--215}.
\newblock


\bibitem[\protect\citeauthoryear{Liu, Wu, Mai, and Lee}{Liu
  et~al\mbox{.}}{2016b}]%
        {liu2016newcache}
\bibfield{author}{\bibinfo{person}{Fangfei Liu}, \bibinfo{person}{Hao Wu},
  \bibinfo{person}{Kenneth Mai}, {and} \bibinfo{person}{Ruby~B Lee}.}
  \bibinfo{year}{2016}\natexlab{b}.
\newblock \showarticletitle{Newcache: Secure cache architecture thwarting cache
  side-channel attacks}.
\newblock \bibinfo{journal}{\emph{IEEE Micro}} \bibinfo{volume}{36},
  \bibinfo{number}{5} (\bibinfo{year}{2016}), \bibinfo{pages}{8--16}.
\newblock


\bibitem[\protect\citeauthoryear{Liu, Yarom, Ge, Heiser, and Lee}{Liu
  et~al\mbox{.}}{2015}]%
        {liu2015last}
\bibfield{author}{\bibinfo{person}{Fangfei Liu}, \bibinfo{person}{Yuval Yarom},
  \bibinfo{person}{Qian Ge}, \bibinfo{person}{Gernot Heiser}, {and}
  \bibinfo{person}{Ruby~B Lee}.} \bibinfo{year}{2015}\natexlab{}.
\newblock \showarticletitle{Last-level cache side-channel attacks are
  practical}. In \bibinfo{booktitle}{\emph{Security and Privacy (SP), 2015 IEEE
  Symposium on}}. IEEE, \bibinfo{pages}{605--622}.
\newblock


\bibitem[\protect\citeauthoryear{Longo, De~Mulder, Page, and Tunstall}{Longo
  et~al\mbox{.}}{2015}]%
        {longo2015soc}
\bibfield{author}{\bibinfo{person}{Jake Longo}, \bibinfo{person}{Elke
  De~Mulder}, \bibinfo{person}{Dan Page}, {and} \bibinfo{person}{Michael
  Tunstall}.} \bibinfo{year}{2015}\natexlab{}.
\newblock \showarticletitle{SoC it to EM: electromagnetic side-channel attacks
  on a complex system-on-chip}. In \bibinfo{booktitle}{\emph{International
  Workshop on Cryptographic Hardware and Embedded Systems}}. Springer,
  \bibinfo{pages}{620--640}.
\newblock


\bibitem[\protect\citeauthoryear{Lyu and Mishra}{Lyu and Mishra}{2018}]%
        {lyu2018survey}
\bibfield{author}{\bibinfo{person}{Yangdi Lyu} {and} \bibinfo{person}{Prabhat
  Mishra}.} \bibinfo{year}{2018}\natexlab{}.
\newblock \showarticletitle{A Survey of Side-Channel Attacks on Caches and
  Countermeasures}.
\newblock \bibinfo{journal}{\emph{Journal of Hardware and Systems Security}}
  \bibinfo{volume}{2}, \bibinfo{number}{1} (\bibinfo{year}{2018}),
  \bibinfo{pages}{33--50}.
\newblock


\bibitem[\protect\citeauthoryear{Maas, Love, Stefanov, Tiwari, Shi, Asanovic,
  Kubiatowicz, and Song}{Maas et~al\mbox{.}}{2013}]%
        {maas2013phantom}
\bibfield{author}{\bibinfo{person}{Martin Maas}, \bibinfo{person}{Eric Love},
  \bibinfo{person}{Emil Stefanov}, \bibinfo{person}{Mohit Tiwari},
  \bibinfo{person}{Elaine Shi}, \bibinfo{person}{Krste Asanovic},
  \bibinfo{person}{John Kubiatowicz}, {and} \bibinfo{person}{Dawn Song}.}
  \bibinfo{year}{2013}\natexlab{}.
\newblock \showarticletitle{Phantom: Practical oblivious computation in a
  secure processor}. In \bibinfo{booktitle}{\emph{Proceedings of the 2013 ACM
  SIGSAC conference on Computer \& communications security}}. ACM,
  \bibinfo{pages}{311--324}.
\newblock


\bibitem[\protect\citeauthoryear{Martin, Demme, and Sethumadhavan}{Martin
  et~al\mbox{.}}{2012}]%
        {martin2012timewarp}
\bibfield{author}{\bibinfo{person}{Robert Martin}, \bibinfo{person}{John
  Demme}, {and} \bibinfo{person}{Simha Sethumadhavan}.}
  \bibinfo{year}{2012}\natexlab{}.
\newblock \showarticletitle{TimeWarp: rethinking timekeeping and performance
  monitoring mechanisms to mitigate side-channel attacks}.
\newblock \bibinfo{journal}{\emph{ACM SIGARCH Computer Architecture News}}
  \bibinfo{volume}{40}, \bibinfo{number}{3} (\bibinfo{year}{2012}),
  \bibinfo{pages}{118--129}.
\newblock


\bibitem[\protect\citeauthoryear{Payer}{Payer}{2016}]%
        {payer2016hexpads}
\bibfield{author}{\bibinfo{person}{Mathias Payer}.}
  \bibinfo{year}{2016}\natexlab{}.
\newblock \showarticletitle{HexPADS: a platform to detect “stealth”
  attacks}. In \bibinfo{booktitle}{\emph{International Symposium on Engineering
  Secure Software and Systems}}. Springer, \bibinfo{pages}{138--154}.
\newblock


\bibitem[\protect\citeauthoryear{Qureshi}{Qureshi}{2018}]%
        {ceaser}
\bibfield{author}{\bibinfo{person}{Moinuddin~K Qureshi}.}
  \bibinfo{year}{2018}\natexlab{}.
\newblock \showarticletitle{CEASER: Mitigating Conflict-Based Cache Attacks via
  Encrypted-Address and Remapping}. In \bibinfo{booktitle}{\emph{2018 51st
  Annual IEEE/ACM International Symposium on Microarchitecture (MICRO)}}. IEEE,
  \bibinfo{pages}{775--787}.
\newblock


\bibitem[\protect\citeauthoryear{Qureshi}{Qureshi}{2019}]%
        {qureshi2019new}
\bibfield{author}{\bibinfo{person}{Moinuddin~K Qureshi}.}
  \bibinfo{year}{2019}\natexlab{}.
\newblock \showarticletitle{New attacks and defense for encrypted-address
  cache}. In \bibinfo{booktitle}{\emph{Proceedings of the 46th International
  Symposium on Computer Architecture}}. ACM, \bibinfo{pages}{360--371}.
\newblock


\bibitem[\protect\citeauthoryear{Rane, Lin, and Tiwari}{Rane
  et~al\mbox{.}}{[n.d.]}]%
        {rane2015raccoon}
\bibfield{author}{\bibinfo{person}{Ashay Rane}, \bibinfo{person}{Calvin Lin},
  {and} \bibinfo{person}{Mohit Tiwari}.} \bibinfo{year}{[n.d.]}\natexlab{}.
\newblock \showarticletitle{Raccoon: Closing Digital Side-Channels through
  Obfuscated Execution.}
\newblock


\bibitem[\protect\citeauthoryear{Sanchez and Kozyrakis}{Sanchez and
  Kozyrakis}{2013}]%
        {sanchez2013zsim}
\bibfield{author}{\bibinfo{person}{Daniel Sanchez} {and}
  \bibinfo{person}{Christos Kozyrakis}.} \bibinfo{year}{2013}\natexlab{}.
\newblock \showarticletitle{ZSim: Fast and accurate microarchitectural
  simulation of thousand-core systems}. In \bibinfo{booktitle}{\emph{ACM
  SIGARCH Computer architecture news}}, Vol.~\bibinfo{volume}{41}. ACM,
  \bibinfo{pages}{475--486}.
\newblock


\bibitem[\protect\citeauthoryear{Seo, Lee, Kim, Shih, Shin, Han, and Kim}{Seo
  et~al\mbox{.}}{2017}]%
        {seo2017sgx}
\bibfield{author}{\bibinfo{person}{Jaebaek Seo}, \bibinfo{person}{Byoungyoung
  Lee}, \bibinfo{person}{Seong~Min Kim}, \bibinfo{person}{Ming-Wei Shih},
  \bibinfo{person}{Insik Shin}, \bibinfo{person}{Dongsu Han}, {and}
  \bibinfo{person}{Taesoo Kim}.} \bibinfo{year}{2017}\natexlab{}.
\newblock \showarticletitle{SGX-Shield: Enabling Address Space Layout
  Randomization for SGX Programs.}
\newblock


\bibitem[\protect\citeauthoryear{Shamir and Tromer}{Shamir and Tromer}{2004}]%
        {shamir2004acoustic}
\bibfield{author}{\bibinfo{person}{Adi Shamir} {and} \bibinfo{person}{Eran
  Tromer}.} \bibinfo{year}{2004}\natexlab{}.
\newblock \showarticletitle{Acoustic cryptanalysis}.
\newblock \bibinfo{journal}{\emph{presentation available from http://www.
  wisdom. weizmann. ac. il/~ tromer}} (\bibinfo{year}{2004}).
\newblock


\bibitem[\protect\citeauthoryear{Sherwood, Perelman, and Calder}{Sherwood
  et~al\mbox{.}}{2001}]%
        {sherwood2001basic}
\bibfield{author}{\bibinfo{person}{Timothy Sherwood}, \bibinfo{person}{Erez
  Perelman}, {and} \bibinfo{person}{Brad Calder}.}
  \bibinfo{year}{2001}\natexlab{}.
\newblock \showarticletitle{Basic block distribution analysis to find periodic
  behavior and simulation points in applications}. In
  \bibinfo{booktitle}{\emph{Parallel Architectures and Compilation Techniques,
  2001. Proceedings. 2001 International Conference on}}. IEEE,
  \bibinfo{pages}{3--14}.
\newblock


\bibitem[\protect\citeauthoryear{Stefanov, Van~Dijk, Shi, Fletcher, Ren, Yu,
  and Devadas}{Stefanov et~al\mbox{.}}{2013}]%
        {stefanov2013path}
\bibfield{author}{\bibinfo{person}{Emil Stefanov}, \bibinfo{person}{Marten
  Van~Dijk}, \bibinfo{person}{Elaine Shi}, \bibinfo{person}{Christopher
  Fletcher}, \bibinfo{person}{Ling Ren}, \bibinfo{person}{Xiangyao Yu}, {and}
  \bibinfo{person}{Srinivas Devadas}.} \bibinfo{year}{2013}\natexlab{}.
\newblock \showarticletitle{Path ORAM: an extremely simple oblivious RAM
  protocol}. In \bibinfo{booktitle}{\emph{Proceedings of the 2013 ACM SIGSAC
  conference on Computer \& communications security}}. ACM,
  \bibinfo{pages}{299--310}.
\newblock


\bibitem[\protect\citeauthoryear{Van~Bulck, Minkin, Weisse, Genkin, Kasikci,
  Piessens, Silberstein, Wenisch, Yarom, and Strackx}{Van~Bulck
  et~al\mbox{.}}{2018}]%
        {vanbulck2018foreshadow}
\bibfield{author}{\bibinfo{person}{Jo Van~Bulck}, \bibinfo{person}{Marina
  Minkin}, \bibinfo{person}{Ofir Weisse}, \bibinfo{person}{Daniel Genkin},
  \bibinfo{person}{Baris Kasikci}, \bibinfo{person}{Frank Piessens},
  \bibinfo{person}{Mark Silberstein}, \bibinfo{person}{Thomas~F. Wenisch},
  \bibinfo{person}{Yuval Yarom}, {and} \bibinfo{person}{Raoul Strackx}.}
  \bibinfo{year}{2018}\natexlab{}.
\newblock \showarticletitle{Foreshadow: Extracting the Keys to the {Intel SGX}
  Kingdom with Transient Out-of-Order Execution}. In
  \bibinfo{booktitle}{\emph{Proceedings of the 27th {USENIX} Security
  Symposium}}. \bibinfo{publisher}{{USENIX} Association}.
\newblock
\newblock
\shownote{See also technical report Forshadow-NG~\cite{weisse2018forshadowNG}
  (\url{https://foreshadowattack.eu}).}


\bibitem[\protect\citeauthoryear{Van~Bulck, Weichbrodt, Kapitza, Piessens, and
  Strackx}{Van~Bulck et~al\mbox{.}}{2017}]%
        {van2017telling}
\bibfield{author}{\bibinfo{person}{Jo Van~Bulck}, \bibinfo{person}{Nico
  Weichbrodt}, \bibinfo{person}{R{\"u}diger Kapitza}, \bibinfo{person}{Frank
  Piessens}, {and} \bibinfo{person}{Raoul Strackx}.}
  \bibinfo{year}{2017}\natexlab{}.
\newblock \showarticletitle{Telling your secrets without page faults: Stealthy
  page table-based attacks on enclaved execution}. In
  \bibinfo{booktitle}{\emph{26th $\{$USENIX$\}$ Security Symposium
  ($\{$USENIX$\}$ Security 17)}}. \bibinfo{pages}{1041--1056}.
\newblock


\bibitem[\protect\citeauthoryear{Wang, Ferraiuolo, Zhang, Myers, and Suh}{Wang
  et~al\mbox{.}}{2016}]%
        {wang2016secdcp}
\bibfield{author}{\bibinfo{person}{Yao Wang}, \bibinfo{person}{Andrew
  Ferraiuolo}, \bibinfo{person}{Danfeng Zhang}, \bibinfo{person}{Andrew~C
  Myers}, {and} \bibinfo{person}{G~Edward Suh}.}
  \bibinfo{year}{2016}\natexlab{}.
\newblock \showarticletitle{SecDCP: secure dynamic cache partitioning for
  efficient timing channel protection}. In \bibinfo{booktitle}{\emph{Design
  Automation Conference (DAC), 2016 53nd ACM/EDAC/IEEE}}. IEEE,
  \bibinfo{pages}{1--6}.
\newblock


\bibitem[\protect\citeauthoryear{Wang and Lee}{Wang and Lee}{2007}]%
        {wang2007new}
\bibfield{author}{\bibinfo{person}{Zhenghong Wang} {and}
  \bibinfo{person}{Ruby~B Lee}.} \bibinfo{year}{2007}\natexlab{}.
\newblock \showarticletitle{New cache designs for thwarting software
  cache-based side channel attacks}. In \bibinfo{booktitle}{\emph{ACM SIGARCH
  Computer Architecture News}}, Vol.~\bibinfo{volume}{35}. ACM,
  \bibinfo{pages}{494--505}.
\newblock


\bibitem[\protect\citeauthoryear{Weisse, Van~Bulck, Minkin, Genkin, Kasikci,
  Piessens, Silberstein, Strackx, Wenisch, and Yarom}{Weisse
  et~al\mbox{.}}{2018}]%
        {weisse2018forshadowNG}
\bibfield{author}{\bibinfo{person}{Ofir Weisse}, \bibinfo{person}{Jo
  Van~Bulck}, \bibinfo{person}{Marina Minkin}, \bibinfo{person}{Daniel Genkin},
  \bibinfo{person}{Baris Kasikci}, \bibinfo{person}{Frank Piessens},
  \bibinfo{person}{Mark Silberstein}, \bibinfo{person}{Raoul Strackx},
  \bibinfo{person}{Thomas~F. Wenisch}, {and} \bibinfo{person}{Yuval Yarom}.}
  \bibinfo{year}{2018}\natexlab{}.
\newblock \showarticletitle{Foreshadow-NG: Breaking the Virtual Memory
  Abstraction with Transient Out-of-Order Execution}.
\newblock \bibinfo{journal}{\emph{Technical report}} (\bibinfo{year}{2018}).
\newblock
\newblock
\shownote{See also USENIX Security paper
  Foreshadow~\cite{vanbulck2018foreshadow}
  (\url{https://foreshadowattack.eu}).}


\bibitem[\protect\citeauthoryear{Wu, Jaleel, Hasenplaugh, Martonosi, Steely~Jr,
  and Emer}{Wu et~al\mbox{.}}{2011}]%
        {wu2011ship}
\bibfield{author}{\bibinfo{person}{Carole-Jean Wu}, \bibinfo{person}{Aamer
  Jaleel}, \bibinfo{person}{Will Hasenplaugh}, \bibinfo{person}{Margaret
  Martonosi}, \bibinfo{person}{Simon~C Steely~Jr}, {and} \bibinfo{person}{Joel
  Emer}.} \bibinfo{year}{2011}\natexlab{}.
\newblock \showarticletitle{SHiP: Signature-based hit predictor for high
  performance caching}. In \bibinfo{booktitle}{\emph{Proceedings of the 44th
  Annual IEEE/ACM International Symposium on Microarchitecture}}. ACM,
  \bibinfo{pages}{430--441}.
\newblock


\bibitem[\protect\citeauthoryear{Yan, Choi, Skarlatos, Morrison, Fletcher, and
  Torrellas}{Yan et~al\mbox{.}}{2018}]%
        {yan2018invisispec}
\bibfield{author}{\bibinfo{person}{Mengjia Yan}, \bibinfo{person}{Jiho Choi},
  \bibinfo{person}{Dimitrios Skarlatos}, \bibinfo{person}{Adam Morrison},
  \bibinfo{person}{Christopher~W Fletcher}, {and} \bibinfo{person}{Josep
  Torrellas}.} \bibinfo{year}{2018}\natexlab{}.
\newblock \showarticletitle{InvisiSpec: Making Speculative Execution Invisible
  in the Cache Hierarchy}. In \bibinfo{booktitle}{\emph{Proceedings of the 51th
  International Symposium on Microarchitecture (MICRO’18)}}.
\newblock


\bibitem[\protect\citeauthoryear{Yan, Gopireddy, Shull, and Torrellas}{Yan
  et~al\mbox{.}}{2017}]%
        {yan2017secure}
\bibfield{author}{\bibinfo{person}{Mengjia Yan}, \bibinfo{person}{Bhargava
  Gopireddy}, \bibinfo{person}{Thomas Shull}, {and} \bibinfo{person}{Josep
  Torrellas}.} \bibinfo{year}{2017}\natexlab{}.
\newblock \showarticletitle{Secure Hierarchy-Aware Cache Replacement Policy
  (SHARP): Defending Against Cache-Based Side Channel Atacks}. In
  \bibinfo{booktitle}{\emph{Proceedings of the 44th Annual International
  Symposium on Computer Architecture}}. ACM, \bibinfo{pages}{347--360}.
\newblock


\bibitem[\protect\citeauthoryear{Yan, Shalabi, and Torrellas}{Yan
  et~al\mbox{.}}{2016}]%
        {yan2016replayconfusion}
\bibfield{author}{\bibinfo{person}{Mengjia Yan}, \bibinfo{person}{Yasser
  Shalabi}, {and} \bibinfo{person}{Josep Torrellas}.}
  \bibinfo{year}{2016}\natexlab{}.
\newblock \showarticletitle{ReplayConfusion: detecting cache-based covert
  channel attacks using record and replay}. In \bibinfo{booktitle}{\emph{The
  49th Annual IEEE/ACM International Symposium on Microarchitecture}}. IEEE
  Press, \bibinfo{pages}{39}.
\newblock


\bibitem[\protect\citeauthoryear{Yarom and Falkner}{Yarom and Falkner}{2014}]%
        {yarom2014flush+}
\bibfield{author}{\bibinfo{person}{Yuval Yarom} {and} \bibinfo{person}{Katrina
  Falkner}.} \bibinfo{year}{2014}\natexlab{}.
\newblock \showarticletitle{FLUSH+ RELOAD: A High Resolution, Low Noise, L3
  Cache Side-Channel Attack.}. In \bibinfo{booktitle}{\emph{USENIX Security
  Symposium}}. \bibinfo{pages}{719--732}.
\newblock


\bibitem[\protect\citeauthoryear{Zhou, Reiter, and Zhang}{Zhou
  et~al\mbox{.}}{2016}]%
        {zhou2016software}
\bibfield{author}{\bibinfo{person}{Ziqiao Zhou}, \bibinfo{person}{Michael~K
  Reiter}, {and} \bibinfo{person}{Yinqian Zhang}.}
  \bibinfo{year}{2016}\natexlab{}.
\newblock \showarticletitle{A software approach to defeating side channels in
  last-level caches}. In \bibinfo{booktitle}{\emph{Proceedings of the 2016 ACM
  SIGSAC Conference on Computer and Communications Security}}. ACM,
  \bibinfo{pages}{871--882}.
\newblock


\end{thebibliography}

\end{document}